\documentclass{article}   

\usepackage{graphicx}  
\usepackage{dcolumn}   
\usepackage{bm}        
\usepackage{amssymb}   

\usepackage{amsmath}   
\usepackage{paralist}  

\usepackage[numbers, sort&compress]{natbib}

\usepackage{color, soul}

\usepackage[labelfont=bf]{caption} 
\usepackage[strict]{changepage} 
\usepackage{textgreek} 

\usepackage{lscape}

\usepackage[strict]{changepage}
\usepackage[labeled]{multibib}
\newcites{S}{Supplementary References}
\usepackage{titling}

\begin{document}
\title{{Non-Reciprocal} Brillouin Scattering Induced Transparency}

\author{JunHwan Kim$^1$, Mark C. Kuzyk$^2$, Kewen Han$^1$, \\Hailin Wang$^2$, Gaurav Bahl$^{1\ast}$\\
	\\
	\footnotesize{$^1$Mechanical Science and Engineering, University of Illinois at Urbana-Champaign,}\\
	\footnotesize{Urbana, Illinois, USA}\\
	\footnotesize{$^2$Department of Physics, University of Oregon}\\
	\footnotesize{Eugene, Oregon, USA}\\
	\footnotesize{$^\ast$To whom correspondence should be addressed; E-mail: bahl@illinois.edu.}
}

\date{}

\maketitle

\begin{abstract}
	

	Electromagnetically induced transparency (EIT)~\cite{Boller:1991if,Hau:1999fp} provides a powerful mechanism for controlling light propagation in a dielectric medium, and for producing slow and fast light. EIT traditionally arises from destructive interference induced by a nonradiative coherence in an atomic system. Stimulated Brillouin scattering (SBS) of light from propagating hypersonic acoustic waves~\cite{PhysRevLett.12.592} has also been used successfully for the generation of slow and fast light~\cite{Song:2005vp,Okawachi:2005cw,Thevenaz:2008vw,Boyd:2009wp}. However, EIT-type processes based on SBS were considered infeasible because of the short coherence lifetime of hypersonic phonons. {Here, we report a new Brillouin scattering induced transparency (BSIT) phenomenon generated by acousto-optic interaction of light with long-lived propagating phonons}~\cite{Bahl:2011cf,Bahl:2012jm}. {We demonstrate that BSIT is uniquely non-reciprocal due to the propagating acoustic phonon wave and accompanying momentum conservation requirement. Using a silica microresonator having naturally occurring forward-SBS phase-matched modal configuration}~\cite{Bahl:2011cf,Bahl:2012jm}{, we show that BSIT enables compact and ultralow-power slow-light generation with delay-bandwidth product comparable to state-of-the-art SBS systems.}
	
\end{abstract}

Stimulated Brillouin scattering (SBS) \cite{PhysRevLett.12.592,PhysRev.137.A1787,Yariv:1965ub} is a fundamental material-level nonlinearity occurring in all states of matter \cite{Boyd} in which two optical fields are coupled to a traveling acoustic wave through photoelastic scattering and electrostriction. The light fields scatter from the periodic photoelastic perturbation generated by the traveling acoustic wave, while simultaneously writing a spatiotemporally beating electrostriction force whose momentum and frequency matches the acoustic wave. Phase matching for SBS is thus defined by both energy and momentum conservation, and is satisfied in back-scattering only by multi-GHz phonon modes in most solids. SBS is frequently used for optical gain \cite{PhysRevLett.12.592,Li:2012bf}, laser linewidth narrowing \cite{Debut:2001tn}, optical phase conjugation \cite{Zeldovich:1972vt}, dynamic gratings \cite{Pant:2013uq}, and even material characterization and microscopy \cite{Montrose:1968uf,Lee:1987fw,Cheng:2006ju,Rich:1972id,Pinnow:1968wr,Scarcelli:2007ha}.
The applications of SBS in superluminal and slow light experiments have been recognized as well \cite{Song:2005vp,Okawachi:2005cw,Thevenaz:2008vw,Boyd:2009wp}.
However, unlike the case of electromagnetically induced transparency (EIT) \cite{Boller:1991if,Hau:1999fp}, the generation of transparency with SBS has never been demonstrated. 
This is because in typical Brillouin scattering pump-probe systems the lifetimes of phonons at multi-GHz frequencies are much shorter than the photon lifetimes \cite{PhysRevLett.12.592,Shin:2013fr,Dainese:2006ki}, effectively disabling the coherent interference of probe Stokes scattering and pump anti-Stokes scattering pathways.
In a forward-scattering SBS system however, this lifetime relationship can be reversed by coupling the light fields through a low-frequency long-lived phonon mode as shown recently in the experimental demonstration of forward-SBS lasing \cite{Bahl:2011cf} and Brillouin cooling \cite{Bahl:2012jm}. 
\begin{figure}[p]
	\begin{adjustwidth}{-1in}{-1in}
		\centering
		\vspace{-100pt}
		\includegraphics{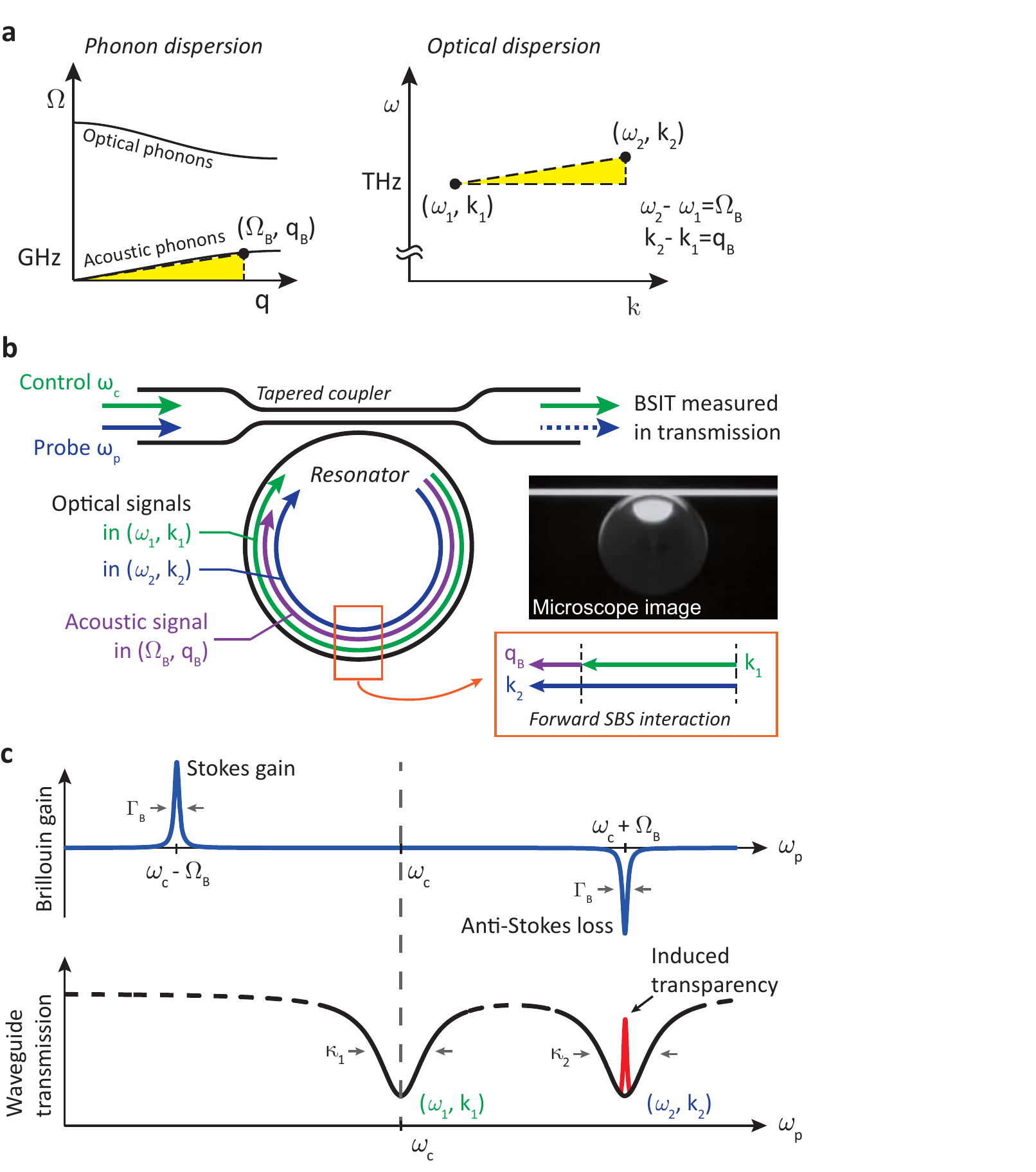}
		\caption{\textbf{Theoretical description of BSIT.} \textbf{a.} Brillouin phase matching requires two optical modes of a system that are separated by the acoustic parameters in $\omega$-k space. \textbf{b.} In a BSIT experiment, control and probe lasers are coupled to phase matched optical whispering gallery modes of a resonator by means of a tapered optical fiber. A forward-SBS opto-acoustic interaction takes place within the resonator. \textbf{c.} The control laser pumps the lower frequency optical mode to allow anti-Stokes scattering while the resonator suppresses Stokes scattering \cite{Bahl:2012jm}. (Bottom) A probe laser scans through the optical resonances, visible as a drop in waveguide transmission \cite{GorodetskyOptimal}. (Top) From traditional SBS theory \cite{Boyd} anti-Stokes absorption is predicted for a probe signal at $\omega_{p}=\omega_{c}+\Omega_{B}$, which intuitively should increase the probe loss. However, when ($\Omega_{B}, q_{B}$) represent a long lived phonon mode, a transparency is observed due to interference between Stokes and anti-Stokes scattering pathways. $\kappa_i$ represent the respective dissipation rates of the two optical modes.}
		\label{theory}
	\end{adjustwidth}
\end{figure}

\begin{figure}[p]
	\begin{adjustwidth}{-1in}{-1in}
		\centering
		\includegraphics[scale=0.8]{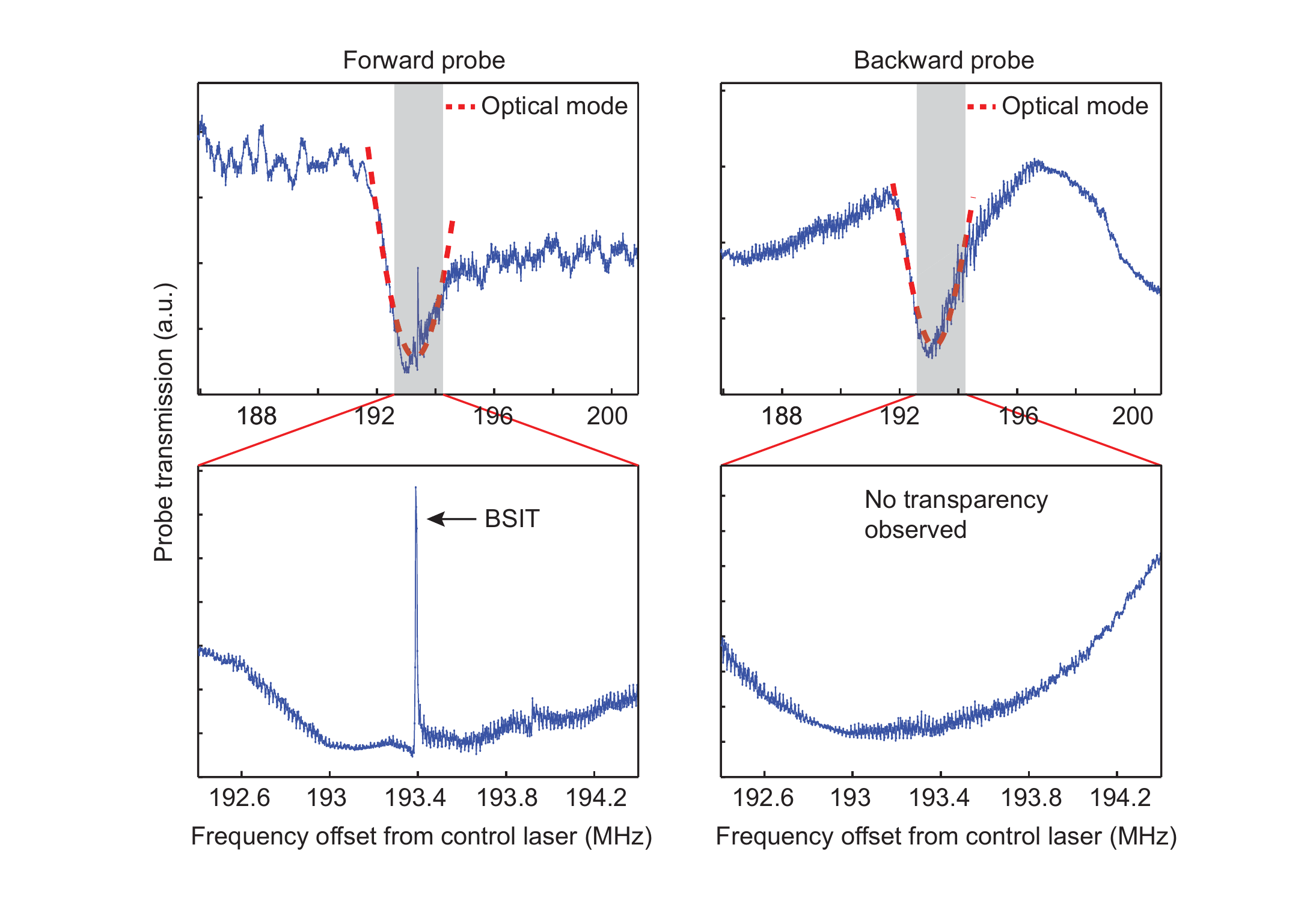}
		\caption{{\textbf{Demonstration of non-reciprocity in BSIT.} (Left) Transparency is induced in the anti-Stokes optical mode for a forward propagating probe signal in a Brillouin coupled system. (Right) There is no transparency observed when the optical probe signal is switched to the backward direction. The background noise and asymmetries are created due to the Fabry-Perot resonances in the differing forward and backward fiber propagation pathways, and due to the transfer functions of the two different electro-optic modulators used for the experiment.}}
		\label{opticalIsolation}
	\end{adjustwidth}
\end{figure}

{Nonlinear optical processes such as EIT need to satisfy the energy-momentum conservation, leading to the phase-matching requirement.  For EIT in an atomic system, the momentum of the spin wave is set by the relevant optical waves, automatically satisfying the phase matching requirement}~\cite{PhysRevA.69.063809,Hau:1999fp,Boller:1991if}. {In comparison, the traveling acoustic wave in Brillouin scattering induced transparency (BSIT) carries a momentum that is intrinsic to the mechanical medium, is discretized by the resonator, and can far exceed the momentum of an optical wave, which leads to special phase matching requirement as will be discussed in detail below.  More importantly, this unique property makes BSIT intrinsically non-reciprocal.  The BSIT vanishes when either the probe or the control reverses its propagation direction (see Fig} \ref{opticalIsolation}). {Such non-reciprocity} \cite{5784283,Poulton2012,Kang.2011} {can be exploited for important applications, including on-chip optical isolators, circulators, and gyroscopes.}

\begin{figure}[p]
	\begin{adjustwidth}{-1in}{-1in}
		\centering
		\includegraphics{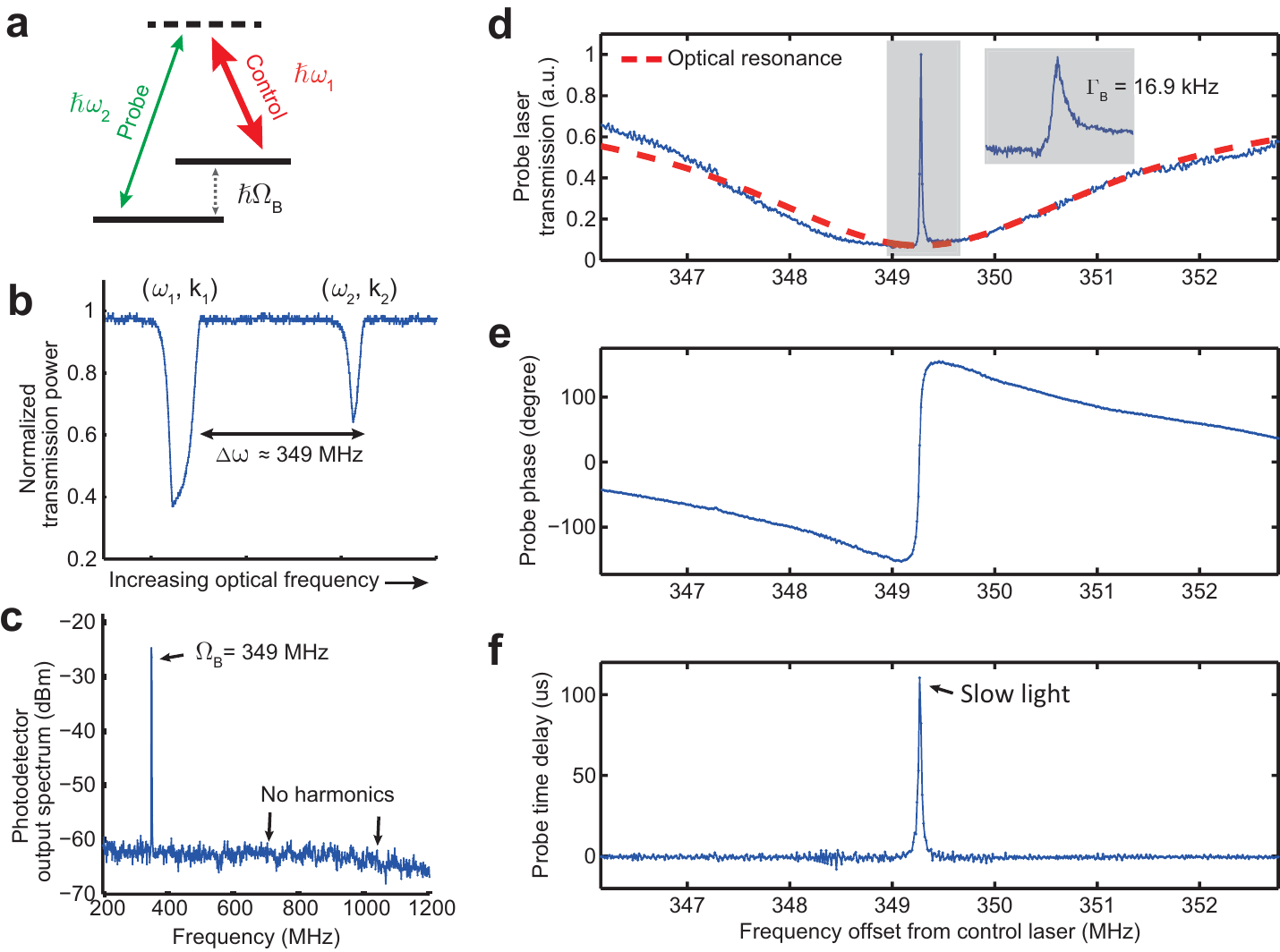}
		\caption{\textbf{Observation of BSIT and slow light.} \textbf{a.} Brillouin scattering analogue of EIT's 3-level lambda system. \textbf{b.} We map resonator optical modes with a swept frequency laser. \textbf{c.} When we pump the ($\omega_2, k_2$) mode with a strong laser source, a Stokes SBS laser is generated \cite{Bahl:2011cf}. When the Stokes scattered and pump light are interfered on a photodetector, a beat note corresponding to their frequency difference, $\Omega_{B}$, is measured. The lack of harmonics of $\Omega_{B}$ indicates that this is not a radiation pressure induced optomechanical oscillation. \textbf{d.} Transparency is observed for a probe on ($\omega_2, k_2$) when a strong pump is placed at ($\omega_1, k_1$). (Solid) Probe laser amplitude response. (Red-dashed) Curve fit to the ($\omega_2, k_2$) optical mode. \textbf{e.} Probe phase response indicates slow light response within the transparency. \textbf{f.} Additional probe group delay $\Delta t = {d\phi}/{d\omega}$ compared to the transmission time without BSIT. 1 mW of laser power is used to generate time delay of 110 {$\mu s$}.}
		\label{antistokes}
	\end{adjustwidth}
\end{figure}

The generation of BSIT requires a 3-mode system composed of two optical modes and one long-lived propagating acoustic mode. Phase-matching of these modes in both frequency space and momentum space is essential as illustrated in Fig.~\ref{theory}a. 
Notably, the momentum requirement is either automatic or unnecessary in other transparency mechanisms~\cite{Boller:1991if,Hau:1999fp,Weis:2010ci,Safavi-Naeini2011,Dong:2013hr}. 
In BSIT, the energies and momenta of these three discrete modes must be matched to satisfy $\omega_2 - \omega_1 = \Omega_B$ and $k_2 - k_1 = q_B$ simultaneously. Here, ($\omega_1, k_1$) are the energy and momentum of the lower energy optical mode, ($\omega_2, k_2$) represent the higher energy optical mode, and ($\Omega_B, q_B$) represent the propagating acoustic mode (Fig.~\ref{theory}a,b). As described above, the coupling between the three modes is mediated by electrostrictive Brillouin scattering.
Previous work has shown that such instances of both forward- and backward-SBS phase matching can occur naturally within a microresonator \cite{GrudininCaF2lasing,Tomes2009,Bahl:2011cf,Bahl:2012jm}.
When momentum matching is assumed, a 3-level lambda system analogy to EIT can also be made in the case of BSIT (Fig.~\ref{antistokes}a).
The system is excited through the addition of a photon at either ($\omega_1, k_1$) or ($\omega_2, k_2$) to a virtual level. The system then radiatively decays to one of the lower levels with the emission of a Stokes or anti-Stokes photon, along with the creation or annihilation of a phonon. The additional conservation of momentum requirement breaks the reciprocity of this system with respect to the optical probe signal \cite{5784283,Poulton2012,Kang.2011}.

{Note that processes analogous to EIT have also been demonstrated recently in optomechanical systems}~\cite{Weis:2010ci,Safavi-Naeini2011}, {in which the optical waves couple to the mechanical motion via radiation pressure.  This optomechanically-induced transparency (OMIT) takes place via stationary breathing vibrational modes, instead of a traveling acoustic wave mode. The forward-SBS phase matching requires co-propagating optical fields with an acoustic field traveling in the same direction} (Fig.~\ref{theory}b). {However, it can be seen that the momentum vector of the chosen acoustic mode does not allow coupling between the two optical fields when the probe field is traveling opposite to the pump. As such, OMIT lacks the special phase matching and non-reciprocity characteristic of BSIT.}

Consider the situation where a strong laser pumps the lower energy optical mode ($\omega_1, k_1$) while a weaker tunable probe beam measures the higher energy optical mode ($\omega_2, k_2$) of the coupled system (Fig.~\ref{antistokes}a). 
When the anti-Stokes probe signal is tuned to a resonator mode, it couples strongly to the resonator and generates a well understood opacity in the waveguide (Fig.~\ref{theory}c,bottom) \cite{GorodetskyOptimal}. 
Further, in the presence of Brillouin phase match with a strong Stokes frequency signal (the control laser) and a long lived phonon mode in the same medium, the anti-Stokes probe is expected to undergo very strong resonant absorption into the Stokes signal (Fig.~\ref{theory}c,top) \cite{Boyd:2009wp}. 
Intuitively, these two cascaded stages of absorption should lead to near-complete removal of the probe light from the system. 
As we show in our system, this intuition breaks down, and instead of strong absorption of the anti-Stokes probe, an interference is generated between the Stokes-directed absorption and anti-Stokes-directed scattering pathways. {This interference of the scattering pathways is analogous to the interference of the optical transitions in EIT. This results in a Brillouin scattering induced transparency at the probe frequency $\omega_{p}=\omega_{c}+\Omega_{B}$ and the probe no longer couples into the resonator optical mode} (Fig.~\ref{theory}c).

In addition to the above intuitive explanation, we rigorously derive the existence of this transparency through the mutually coupled electromagnetic and acoustic wave equations in the Supplement. 
We adopt the formalism for triply-resonant Brillouin scattering developed by Agarwal and Jha in ref.~\cite{PhysRevA.88.013815} to describe our system.
The main result is that the probe field transfer function $a_{2}$/$f_{p}$ in this system is provided by the equations:
\begin{eqnarray}
\frac{a_{2}}{f_{p}} & = & \frac{\Gamma_{B}+i\Delta_{B}}{\left(\kappa+i\Delta_{2}\right)(\Gamma_{B}+i\Delta_{B})+|\beta|^{2}|a_{1}|^{2}} \label{scattered}\\
\Delta_{B} & = & \Omega_{B}-(\omega_{p}-\omega_{c})\\
\Delta_{2} & = & \omega_{2}-\omega_{p}
\end{eqnarray}

where $a_{1}$ and $a_{2}$ are intracavity control and probe light fields respectively, $f_{p}$ is the source field associated with the coupled probe laser, $\kappa$ is optical loss rate, $\Gamma_{B}$ is acoustic loss rate, $\beta$ is coupling coefficient accounting for modal overlap and Brillouin gain in the material, $\Delta_{B}$ and $\Delta_{2}$ are detuning parameters,  $\Omega_{B}$ is Brillouin acoustic mode frequency, $\omega_{2}$ is the anti-Stokes optical mode center frequency, and $\omega_{p}$ and $\omega_{c}$ are probe and control laser frequencies respectively.
{We assume that $\kappa_1=\kappa_2=\kappa$ for mathematical convenience although the optical loss rates are not necessarily identical in practice.}
Here, it can be seen that the acoustic loss rate $\Gamma_{B}$ must be lower than the optical loss rate $\kappa$ in order to observe transparency. In the opposite case where $\kappa \ll \Gamma_{B}$ as in most SBS experiments, the coupling rate $|\beta|^2|a_1|^2$ in eq. \ref{scattered} becomes negligible and no transparency is generated.

\vspace{12pt}

Our experimental setup consists of an ultra-high-Q silica microsphere resonator with approximate diameter of 150 $\mu$m that is pumped by a 1550 nm tunable diode laser by means of a tapered optical fiber as shown in Fig.~\ref{theory}b. For technical convenience, an electro optic modulator (EOM) is used to generate the probe signal at a fixed offset from the control laser (Supplement). A high-speed photodetector monitors the transmitted optical signals at the far end of the tapered fiber, and performs a heterodyned measurement of the probe signal response by means of the beat note between the probe laser and the fixed control laser. We use a network analyzer to generate the EOM input signal (modulation frequency) for the probe laser offset, and to measure the magnitude and phase response of the probe. This experiment is carried out in a room temperature and atmospheric pressure environment.

\begin{figure}[p]
	\begin{adjustwidth}{-1in}{-1in}
		\centering
		\includegraphics{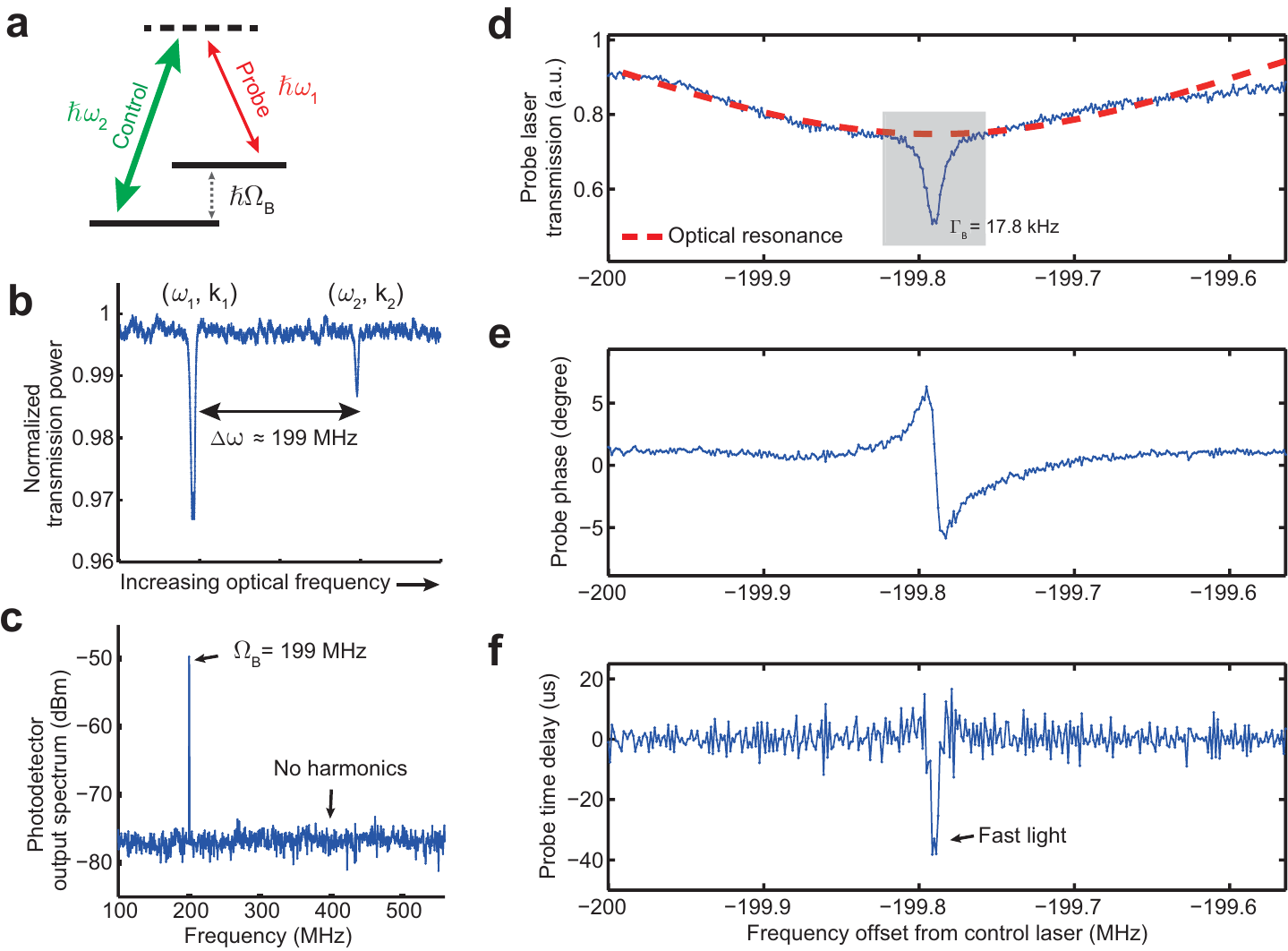}
		\caption{\textbf{Observation of Brillouin scattering induced opacity and fast light.} \textbf{a.} The control laser pumps the ($\omega_2, k_2$) optical mode while the probe measures the lower frequency mode, ($\omega_1, k_1$). \textbf{b.} We map the optical modes with a swept frequency laser. \textbf{c.} When a laser pumps ($\omega_2, k_2$) above threshold (as in Fig.~\ref{antistokes}), forward-SBS lasing results \textbf{d.} In the configuration shown in subfigure (a), the probe experiences an opacity at $\Omega_{B}$ offset from the control laser. (Solid) Amplitude response of the probe laser. (Red-dashed) Curve fit to the ($\omega_1, k_1$) optical mode. \textbf{e.} Negative slope in the probe phase corresponds to a fast light response within the opacity region. \textbf{f.} Probe time advancement compared to the transmission time without Brillouin scattering induced opacity. Time advancement of 35 {$\mu s$} is demonstrated using 380 $\mu$W of control laser power.}
		\label{stokes}
	\end{adjustwidth}
\end{figure}

Experimentally, a system of Brillouin phase-matched optical and acoustic modes is identified through forward-SBS lasing by pumping the higher frequency optical mode above the lasing threshold \cite{Bahl:2011cf}. 
The electronic signature of such phase matching is the generation of a single beat note at $\Omega_{B}$ at the output of the wide bandwidth photodetector, whose lack of harmonics rule out the well known radiation pressure parametric instability as shown in Fig.~\ref{antistokes}c and \ref{stokes}c \cite{PhysRevLett.95.033901}.
When the control laser is parked at the lower optical resonance, it leads to destructive interference for a probe beam on the higher resonance and generates a transparency window (Fig.~\ref{antistokes}a,d).
The measured amplitude response of the probe laser in Fig.~\ref{antistokes}d shows the relatively broad 4.4 MHz wide anti-Stokes optical mode with optical quality factor of $4.4 \times 10^7$. A very sharp transparency feature at 349.3 MHz offset from the control laser is observed in both amplitude and phase responses of the probe (Fig.~\ref{antistokes}d,e). This 349.3 MHz frequency corresponds to a whispering-gallery acoustic wave mode \cite{Bahl:2011cf} of the resonator with azimuthal mode number of 48 and phonon lifetime of 59.2 $\mu s$. The rapidly changing phase response feature with a positive slope corresponds to a very low group velocity for the optical probe i.e. ``slow light''\cite{Boyd:2002ud, Safavi-Naeini2011}. We calculate the group delay, $\Delta t$, using the relationship $\Delta t = {d\phi}/{d\omega}$. Positive $\Delta t$ represents an optical delay and a negative value represents an optical advancement.
Here, 1 mW of control laser power is used to generate 110 $\mu$s time delay for the probe (Fig.~\ref{antistokes}f).

We note that this observation of slow light with an anti-Stokes probe signal appears to stand in opposition to the previously demonstrated SBS-induced fast light with the chosen pump-probe configuration \cite{Okawachi:2005cw,Thevenaz:2008vw,Boyd:2002ud,Boyd:2009wp,Pant2012}. This is because the probe phase response in our experiment is measured in the waveguide, whereas the SBS interaction takes place within the resonator. As we detail in the Supplement, the probe phase response within the resonator precisely follows the expected fast light response from SBS systems. Since the waveguide transmission is a result of the interference of the exiting probe light from the resonator with the externally supplied probe signal, the measured response exhibits a slow light characteristic.

On the other hand, if the control optical signal is tuned to the higher energy optical resonance, it creates a region of greater opacity for a probe on the lower energy optical mode and a``fast light'' dispersion feature (Fig.~\ref{stokes}a,d).
Due to thermal effects on the optical resonance frequencies, we found it technically more convenient to demonstrate opacity and fast light with a different forward-SBS triplet on the same resonator where $\Omega_{B}$= 199.8 MHz (phonon lifetime of 56.2 $\mu s$). The amplitude response of the probe laser in Fig.~\ref{stokes}d shows the 0.4 MHz wide Stokes optical mode with optical quality factor of $4.9 \times 10^8$. An opacity with a 17.8 kHz linewidth is induced at 199.8 MHz offset from the pump. Again, based on the rapidly varying phase response with a  negative slope, we estimate additional 35 $\mu$s fast light time advancement using 380 $\mu$W of control laser power (Fig.~\ref{stokes}f).

{We note that for the induced absorption in BSIT (or even OMIT) to occur, the driving or pump field leads to an amplification instead of damping of the mechanical mode. In this regard, induced absorption in BSIT is related to optical parametric amplification in a resonator (below the threshold).  In this context, the induced absorption in BSIT (or OMIT) differs from the effect of Electromagnetically Induced Absorption (EIA)}~\cite{PhysRevA.59.4732} {in atomic systems.}

\vspace{12pt}

{We now discuss the achievable time delay and bandwidth for slow light through BSIT. The acoustic dissipation rate $\Gamma_B$ is the primary determinant for the bandwidth over which BSIT occurs and where slow light can be obtained. In such forward-SBS systems, $\Gamma_B$ can be increased or decreased through the control laser power as previously shown in the experiment on Brillouin cooling}~\cite{Bahl:2012jm}. {In contrast, for linear systems based on backward SBS, the phonon loss rate is orders of magnitude higher than the photon loss rate}~\cite{Boyd}, {and thus the acoustic dissipation rate remains unaffected for all practical control laser powers. Specifically in the case of BSIT and forward SBS cooling}~\cite{Bahl:2012jm} {in resonators, $\Gamma_B = \Gamma_i + \left( | \beta |^2 |a_1 |^2 \kappa \right) / \left( \kappa^2 + \delta^2 \right)$ describes the relationship of the acoustic dissipation with respect to the control laser field $a_1$}~\cite{PhysRevA.88.013815}. {Here, $\Gamma_i$ is the intrinsic acoustic dissipation rate for the unperturbed traveling-wave acoustic mode, while $\delta$ is the detuning of the transparency with respect to the anti-Stokes optical mode ($\delta = \omega_2 - \omega_1 - \Omega_B$ as described in the Supplement).} 
{The time delay achievable in BSIT is expressed through the relation $\Delta t = 2 / \Gamma_B$ derived precisely as in other opto-acoustic transparency mechanisms}~\cite{Weis:2010ci,Safavi-Naeini2011}. {This expression underscores the delay vs bandwidth tradeoff that must be made and that cannot be overcome in forward-SBS or optomechanical systems.}

{The delay-bandwidth product generally represents the storage capacity of SBS slow light systems}~\cite{4139648} {and the penalty for higher delay times is a reduced bandwidth} \cite{Song:2005vp,Okawachi:2005cw,Thevenaz:2008vw,4139648} {even when spectrally broadened control lasers are employed}~\cite{Herraez2006}. {Increased control laser power or a longer waveguide can be used in linear SBS systems to improve the opto-acoustic interaction gain and to compensate for this penalty}~\cite{Herraez2006,Song:2005vp,Okawachi:2005cw,Thevenaz:2008vw,4139648}. {Lower input power implies lower Brillouin gain and thus lower delay-bandwidth. According to the equations above, however, BSIT slow light behaves the opposite way with respect to increasing input power.}

The delay-bandwidth product reported here, 0.63 MHz-$\mu$s for fast light and 1.87 MHz-$\mu$s for slow light, is on par with previous SBS based demonstrations \cite{Pant2012,Okawachi:2005cw,Song:2005vp,6582552,4139648}. We provide a complete comparison against prior literature in the Supplement. We also consider the space and power budgets used to generate the optical delay. 
{One can account for the device footprint i.e. opto-acoustic interaction length (waveguide length in linear SBS systems, circumference in circular resonators) and for the control power input to the device (waveguide or resonator) by normalizing the reported delay-bandwidth product against these two engineering parameters. 
	In our resonator BSIT demonstration, the delay-bandwidth per-laser-power per-interaction-length is about $4.0 \times 10^6$ MHz-$\mu s$ $W^{-1}m^{-1}$, which is about 5 orders-of-magnitude higher than the next highest reported value in any SBS system (44 MHz-$\mu s$ $W^{-1}m^{-1}$) }\cite{Pant2012}. {
	This ultralow power and compact microresonator based slow light demonstration could be thus adaptable for on chip Brillouin systems} \cite{Shin:2013fr,Li:2013bl,Poulton2012} {without needing lengthy waveguides or kilometers of fiber.}

\vspace{12pt}

{We have experimentally validated the non-reciprocity intrinsic to BSIT in} Fig.~\ref{opticalIsolation}. {Here, we independently probed the forward and backward directions in a BSIT system for the same fixed control laser (see Supplement for experimental setup).} Fig.~\ref{opticalIsolation} {shows the same anti-Stokes optical mode in both forward and backward directions, which we validated by modifying the optical coupling, polarization, and laser detuning. Most notably, while a narrowband BSIT is observed within the optical mode in the forward direction, it is entirely absent for a backward probe. This is the first experimental demonstration where non-reciprocal transmission has been observed in an optomechanical system. BSIT thus offers a non-magnetic alternative to Faraday effect based non-reciprocal devices in a small footprint that is appealing for on-chip applications.}

\vspace{12pt}

\begin{figure}[p]
	\begin{adjustwidth}{-1.5in}{-1.5in}
		\centering
		\includegraphics[width=1.6\textwidth]
		{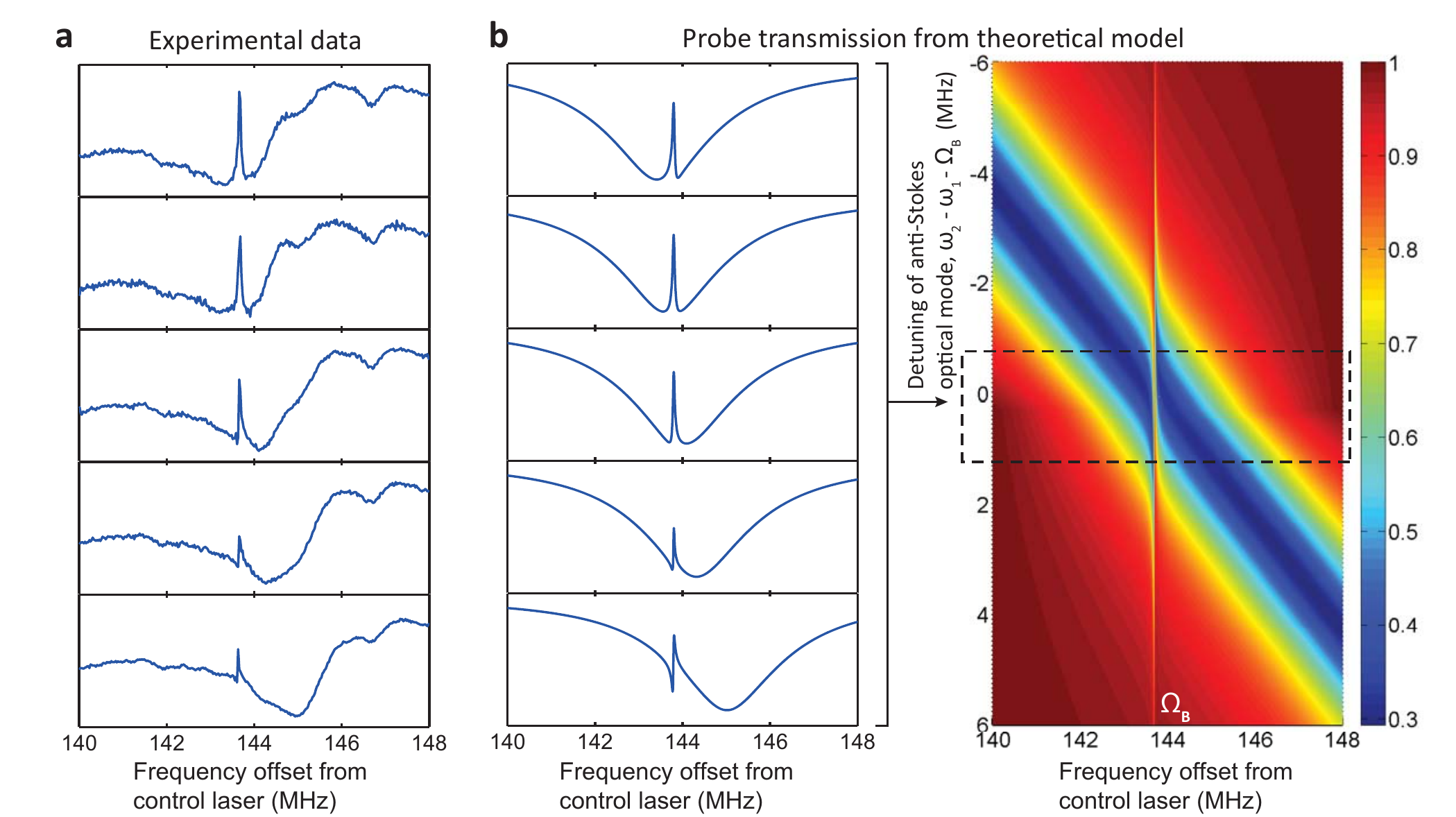}
		
		\caption{ \textbf{a.} Tuning the control laser frequency $\omega_{c}$ causes a relative detuning of the anti-Stokes optical mode while the transparency remains at a fixed frequency offset $\Omega_{B}$. In this example, a SBS triplet with $\Omega_{B}$= 143.7 MHz is used. \textbf{b.} Theoretical modeling through eq.~\ref{scattered} shows the stationary transparency at $\Omega_B$ while the anti-Stokes optical mode is tuned through the control laser frequency. This is consistent with experimental result.}
		\label{moving}
	\end{adjustwidth}
\end{figure}

Although phase matching imposes strict constraints on the optical signal frequencies in the BSIT process, the transparency can be tuned by either slightly modifying the pump frequency within its optical mode or by thermally tuning the optical modes themselves (Fig.~\ref{moving}). Furthermore, the transparency depth and width can be controlled \cite{Bahl:2012jm} through the control laser power represented as $|a_1|^2$ in eq.~\ref{scattered}. Such frequency tunability and the ability to switch the transparency on and off are desirable in several applications \cite{5784283,Poulton2012,Kang.2011}.

\vspace{12pt}

{There is an important difference between EIT processes in an atomic system}~\cite{RevModPhys.77.633} {and EIT-related processes involving a single mechanical oscillator. The latter, including OMIT and BSIT, is essentially a coupled oscillator phenomenon arising from the coherent coupling between the mechanical oscillator and an optical probe field (in the undepleted pump approximation). Spatially extended excitations similar to dark-state polaritons in atomic media}~\cite{Fleischhauer2000} {cannot be formed in a single mechanical oscillator. A direct consequence of this is that only a single mode field can be stored in the mechanical oscillator as discussed by Fiore et. al.}~\cite{PhysRevLett.107.133601}. {An array of oscillators is required for the storage of the spatial-temporal profile of the optical field. This is also similar to the pure optical analog of EIT processes}~\cite{PhysRevLett.92.083901}.

Electrostrictive effects and Brillouin scattering are ubiquitous in all states of matter \cite{Boyd}, including liquids, gases, and plasmas, implying that BSIT can be applied for optical switching in a broad range of experimental platforms potentially even including linear systems \cite{Dainese:2006ki,Shin:2013fr,Poulton2012}. {In contrast to EIT-like phenomena including OMIT}~\cite{Weis:2010ci,Safavi-Naeini2011}, {BSIT requires a stringent momentum conservation, resulting in our demonstration that back-propagating probe signals do not experience the transparency}. Such non-reciprocity can be exploited in {magnet-free on-chip optical isolators, gyroscopes, and circulators}. Moreover, while this demonstration is in a silica sphere, geometry engineering can be used to obtain enhanced SBS effects \cite{Rakich:2012et}, much larger electrostriction coefficients can be obtained in chalcogenide glasses \cite{Poulton2012}, and the sign of electrostrictive coupling can even be negative \cite{turik2006negative}. Such flexibility could lead to new and surprising applications and engineered nonlinearities in the future.

\section*{Acknowledgements}
Funding for this research was provided through a University of Illinois Startup Grant, Office of the Vice Chancellor for Research Research Board Grant, the National Science Foundation, and the Air Force Office for Scientific Research. MCK and HW are supported in part by NSF grant No. 1205544. JHK and GB are supported in part by NSF grant No. 1408539 and AFOSR grant FA9550-14-1-0217. We would like to acknowledge stimulating discussions and guidance from Dr. Kimani Toussaint, Dr. Lucas Wagner, Dr. Rashid Bashir, Dr. Logan Liu, Dr. Lynford Goddard, and Dr. Peter Dragic.

\def\url#1{}
\bibliographystyle{myIEEEtran}
\bibliography{BSITbib}

\begin{thebibliography}{10}
\providecommand{\url}[1]{#1}
\csname url@samestyle\endcsname
\providecommand{\newblock}{\relax}
\providecommand{\bibinfo}[2]{#2}
\providecommand{\BIBentrySTDinterwordspacing}{\spaceskip=0pt\relax}
\providecommand{\BIBentryALTinterwordstretchfactor}{4}
\providecommand{\BIBentryALTinterwordspacing}{\spaceskip=\fontdimen2\font plus
\BIBentryALTinterwordstretchfactor\fontdimen3\font minus
  \fontdimen4\font\relax}
\providecommand{\BIBforeignlanguage}[2]{{%
\expandafter\ifx\csname l@#1\endcsname\relax
\typeout{** WARNING: IEEEtran.bst: No hyphenation pattern has been}%
\typeout{** loaded for the language `#1'. Using the pattern for}%
\typeout{** the default language instead.}%
\else
\language=\csname l@#1\endcsname
\fi
#2}}
\providecommand{\BIBdecl}{\relax}
\BIBdecl

\bibitem{Boller:1991if}
K.~J. Boller, A.~Imamolu, and S.~Harris, ``{Observation of electromagnetically
  induced transparency},'' \emph{Phys. Rev. Lett.}, vol.~66, no.~20, pp.
  2593--2596, May 1991.

\bibitem{Hau:1999fp}
L.~V. Hau, S.~E. Harris, Z.~Dutton, and C.~H. Behroozi, ``{Light speed
  reduction to 17 metres per second in an ultracold atomic gas},''
  \emph{Nature}, vol. 397, no. 6720, pp. 594--598, 1999.

\bibitem{PhysRevLett.12.592}
R.~Y. Chiao, C.~H. Townes, and B.~P. Stoicheff, ``{Stimulated Brillouin
  Scattering and Coherent Generation of Intense Hypersonic Waves},''
  \emph{Phys. Rev. Lett.}, vol.~12, no.~21, pp. 592--595, May 1964.

\bibitem{Song:2005vp}
K.~Y. Song, M.~G. Herr{\'a}ez, and L.~Th{\'e}venaz, ``{Observation of pulse
  delaying and advancement in optical fibers using stimulated Brillouin
  scattering},'' \emph{Opt. Express}, vol.~13, no.~1, pp. 82--88, 2005.

\bibitem{Okawachi:2005cw}
Y.~Okawachi, M.~Bigelow, J.~Sharping, Z.~Zhu, A.~Schweinsberg, D.~Gauthier,
  R.~Boyd, and A.~Gaeta, ``{Tunable All-Optical Delays via Brillouin Slow Light
  in an Optical Fiber},'' \emph{Physical Review Letters}, vol.~94, no.~15, p.
  153902, Apr. 2005.

\bibitem{Thevenaz:2008vw}
L.~Th{\'e}venaz, ``{Slow and fast light in optical fibres},'' \emph{Nature
  Photonics}, vol.~2, no.~8, pp. 474--481, 2008.

\bibitem{Boyd:2009wp}
R.~W. Boyd and D.~J. Gauthier, ``{Controlling the velocity of light pulses},''
  \emph{Science}, vol. 326, no. 5956, pp. 1074--1077, 2009.

\bibitem{Bahl:2011cf}
G.~Bahl, J.~Zehnpfennig, M.~Tomes, and T.~Carmon, ``{Stimulated optomechanical
  excitation of surface acoustic waves in a microdevice},'' \emph{Nature
  Communications}, vol.~2, p. 403, Jul. 2011.

\bibitem{Bahl:2012jm}
G.~Bahl, M.~Tomes, F.~Marquardt, and T.~Carmon, ``{Observation of spontaneous
  Brillouin cooling},'' \emph{Nature Physics}, vol.~8, no.~3, pp. 203--207,
  Mar. 2012.

\bibitem{PhysRev.137.A1787}
Y.~R. Shen and N.~Bloembergen, ``{Theory of Stimulated Brillouin and Raman
  Scattering},'' \emph{Phys. Rev.}, vol. 137, no.~6A, pp. A1787--A1805, Mar
  1965.

\bibitem{Yariv:1965ub}
A.~Yariv, ``{Quantum theory for parametric interactions of light and
  hypersound},'' \emph{Quantum Electronics, IEEE Journal of}, vol.~1, no.~1,
  pp. 28--36, 1965.

\bibitem{Boyd}
R.~W. Boyd, \emph{{Nonlinear Optics}}, 3rd~ed.\hskip 1em plus 0.5em minus
  0.4em\relax Elsevier, 2008, {Chapter 9}.

\bibitem{Li:2012bf}
J.~Li, H.~Lee, T.~Chen, and K.~J. Vahala, ``{Characterization of a high
  coherence, Brillouin microcavity laser on silicon.}'' \emph{Optics Express},
  vol.~20, no.~18, pp. 20\,170--20\,180, Aug. 2012.

\bibitem{Debut:2001tn}
A.~Debut, S.~Randoux, and J.~Zemmouri, ``{Experimental and theoretical study of
  linewidth narrowing in Brillouin fiber ring lasers},'' \emph{JOSA B},
  vol.~18, no.~4, p. 556, 2001.

\bibitem{Zeldovich:1972vt}
B.~Y. Zel'dovich, V.~I. Popocivhec, R.~V, V, and F.~S. Faisullov, ``{Connection
  between the wave fronts of the reflected and exciting light in stumulated
  Mandelshtam-Brillouin scattering},'' \emph{JETP Letters}, vol.~15, no. 109,
  Aug. 1972.

\bibitem{Pant:2013uq}
R.~Pant, E.~Li, C.~G. Poulton, D.-Y. Choi, S.~Madden, B.~Luther-Davies, and
  B.~J. Eggleton, ``{Observation of Brillouin dynamic grating in a photonic
  chip},'' \emph{Optics Letters}, vol.~38, no.~3, pp. 305--307, 2013.

\bibitem{Montrose:1968uf}
C.~J. Montrose, V.~A. Solovyev, and T.~A. Litovitz, ``{Brillouin scattering and
  relaxation in liquids},'' \emph{The Journal of the Acoustical Society of
  America}, vol.~43, p. 117, 1968.

\bibitem{Lee:1987fw}
S.~A. Lee, S.~M. Lindsay, J.~W. Powell, T.~Weidlich, N.~J. Tao, G.~D. Lewen,
  and A.~Rupprecht, ``{A Brillouin scattering study of the hydration of Li- and
  Na-DNA films},'' \emph{Biopolymers}, vol.~26, no.~10, pp. 1637--1665, Oct.
  1987.

\bibitem{Cheng:2006ju}
W.~Cheng, J.~Wang, U.~Jonas, G.~Fytas, and N.~Stefanou, ``{Observation and
  tuning of hypersonic bandgaps in colloidal crystals},'' \emph{Nature
  Materials}, vol.~5, no.~10, pp. 830--836, Sep. 2006.

\bibitem{Rich:1972id}
T.~C. Rich and D.~A. Pinnow, ``{Total Optical Attenuation in Bulk Fused
  Silica},'' \emph{Applied Physics Letters}, vol.~20, no.~7, pp. 264--266,
  1972.

\bibitem{Pinnow:1968wr}
D.~A. Pinnow, S.~J. Candau, J.~T. LaMacchia, and T.~A. Litovitz, ``{Brillouin
  scattering: viscoelastic measurements in liquids},'' \emph{The Journal of the
  Acoustical Society of America}, vol.~43, p. 131, 1968.

\bibitem{Scarcelli:2007ha}
G.~Scarcelli and S.~H. Yun, ``{Confocal Brillouin microscopy for
  three-dimensional mechanical imaging},'' \emph{Nature Photonics}, vol.~2,
  no.~1, pp. 39--43, Dec. 2007.

\bibitem{Shin:2013fr}
H.~Shin, W.~Qiu, R.~Jarecki, J.~A. Cox, R.~H. Olsson, III, A.~Starbuck,
  Z.~Wang, and P.~T. Rakich, ``{Tailorable stimulated Brillouin scattering in
  nanoscale silicon waveguides},'' \emph{Nature Communications}, vol.~4, 2013.

\bibitem{Dainese:2006ki}
P.~Dainese, P.~S.~J. Russell, N.~Joly, J.~C. Knight, G.~S. Wiederhecker, H.~L.
  Fragnito, V.~Laude, and A.~Khelif, ``{Stimulated Brillouin scattering from
  multi-GHz-guided acoustic phonons in nanostructured photonic crystal
  fibres},'' \emph{Nature Physics}, vol.~2, no.~6, pp. 388--392, May 2006.

\bibitem{GorodetskyOptimal}
M.~Gorodetsky and V.~S. Ilchenko, ``{Optical microsphere resonators: optimal
  coupling to high-Q whispering-gallery modes},'' \emph{Journal of the Optical
  Society of America B}, vol.~16, no.~1, pp. 147--154, 1999.

\bibitem{PhysRevA.69.063809}
\BIBentryALTinterwordspacing
P.~Arve, P.~J\"anes, and L.~Thyl\'en, ``Propagation of two-dimensional pulses
  in electromagnetically induced transparency media,'' \emph{Phys. Rev. A},
  vol.~69, p. 063809, Jun 2004.
  \url{http://link.aps.org/doi/10.1103/PhysRevA.69.063809}
\BIBentrySTDinterwordspacing

\bibitem{5784283}
X.~Huang and S.~Fan, ``Complete all-optical silica fiber isolator via
  stimulated brillouin scattering,'' \emph{Lightwave Technology, Journal of},
  vol.~29, no.~15, pp. 2267--2275, Aug 2011.

\bibitem{Poulton2012}
C.~G. Poulton, R.~Pant, A.~Byrnes, S.~Fan, M.~J. Steel, and B.~J. Eggleton,
  ``Design for broadband on-chip isolator using stimulated brillouin scattering
  in dispersion-engineered chalcogenide waveguides,'' \emph{Opt. Express},
  vol.~20, no.~19, pp. 21\,235--21\,246, Sep 2012.

\bibitem{Kang.2011}
\BIBentryALTinterwordspacing
M.~S. Kang, A.~Butsch, and P.~S.~J. Russell, ``Reconfigurable light-driven
  opto-acoustic isolators in photonic crystal fibre,'' \emph{Nat Photon},
  vol.~5, no.~9, pp. 549--553, Sep. 2011.
  \url{http://dx.doi.org/10.1038/nphoton.2011.180}
\BIBentrySTDinterwordspacing

\bibitem{Weis:2010ci}
S.~Weis, R.~Rivi{\`e}re, S.~Del{\'e}glise, E.~Gavartin, O.~Arcizet,
  A.~Schliesser, and T.~J. Kippenberg, ``{Optomechanically induced
  transparency.}'' \emph{Science}, vol. 330, no. 6010, pp. 1520--1523, Dec.
  2010.

\bibitem{Safavi-Naeini2011}
A.~H. Safavi-Naeini, T.~P.~M. Alegre, J.~Chan, M.~Eichenfield, M.~Winger,
  Q.~Lin, J.~T. Hill, D.~E. Chang, and O.~Painter, ``Electromagnetically
  induced transparency and slow light with optomechanics,'' \emph{Nature}, vol.
  472, no. 7341, pp. 69--73, Apr. 2011.

\bibitem{Dong:2013hr}
C.~Dong, V.~Fiore, M.~C. Kuzyk, and H.~Wang, ``{Transient optomechanically
  induced transparency in a silica microsphere},'' \emph{Physical Review A},
  vol.~87, no.~5, p. 055802, May 2013.

\bibitem{GrudininCaF2lasing}
I.~S. Grudinin, A.~B. Matsko, and L.~Maleki, ``Brillouin lasing with a
  {$CaF_{2}$} whispering gallery mode resonator,'' \emph{Phys. Rev. Lett.},
  vol. 102, no.~4, p. 043902, Jan 2009.

\bibitem{Tomes2009}
M.~Tomes and T.~Carmon, ``Photonic micro-electromechanical systems vibrating at
  {X-band} ({11-GHz}) rates,'' \emph{Phys. Rev. Lett.}, vol. 102, no.~11, p.
  113601, March 2009.

\bibitem{PhysRevA.88.013815}
G.~S. Agarwal and S.~S. Jha, ``Multimode phonon cooling via three-wave
  parametric interactions with optical fields,'' \emph{Phys. Rev. A}, vol.~88,
  p. 013815, Jul 2013.

\bibitem{PhysRevLett.95.033901}
T.~J. Kippenberg, H.~Rokhsari, T.~Carmon, A.~Scherer, and K.~J. Vahala,
  ``Analysis of radiation-pressure induced mechanical oscillation of an optical
  microcavity,'' \emph{Phys. Rev. Lett.}, vol.~95, no.~3, p. 033901, Jul 2005.

\bibitem{Boyd:2002ud}
R.~W. Boyd and D.~J. Gauthier, ``{``Slow'' and ``fast'' light},''
  \emph{Progress in Optics}, vol.~43, pp. 497--530, 2002.

\bibitem{Pant2012}
R.~Pant, A.~Byrnes, C.~G. Poulton, E.~Li, D.-Y. Choi, S.~Madden,
  B.~Luther-Davies, and B.~J. Eggleton, ``Photonic-chip-based tunable slow and
  fast light via stimulated brillouin scattering,'' \emph{Opt. Lett.}, vol.~37,
  no.~5, pp. 969--971, Mar 2012.

\bibitem{PhysRevA.59.4732}
\BIBentryALTinterwordspacing
A.~Lezama, S.~Barreiro, and A.~M. Akulshin, ``Electromagnetically induced
  absorption,'' \emph{Phys. Rev. A}, vol.~59, pp. 4732--4735, Jun 1999.
  \url{http://link.aps.org/doi/10.1103/PhysRevA.59.4732}
\BIBentrySTDinterwordspacing

\bibitem{4139648}
L.~Yi, L.~Zhan, W.~Hu, and Y.~Xia, ``Delay of broadband signals using slow
  light in stimulated brillouin scattering with phase-modulated pump,''
  \emph{Photonics Technology Letters, IEEE}, vol.~19, no.~8, pp. 619--621,
  April 2007.

\bibitem{Herraez2006}
\BIBentryALTinterwordspacing
M.~G. Herr\'{a}ez, K.~Y. Song, and L.~Th\'{e}venaz, ``Arbitrary-bandwidth
  brillouin slow light in optical fibers,'' \emph{Opt. Express}, vol.~14,
  no.~4, pp. 1395--1400, Feb 2006.
  \url{http://www.opticsexpress.org/abstract.cfm?URI=oe-14-4-1395}
\BIBentrySTDinterwordspacing

\bibitem{6582552}
H.~Ju, L.~Ren, X.~Lin, J.~Liang, and C.~Ma, ``Wide-range continuously-tunable
  slow-light delay line based on stimulated brillouin scattering,''
  \emph{Photonics Technology Letters, IEEE}, vol.~25, no.~19, pp. 1920--1923,
  Oct 2013.

\bibitem{Li:2013bl}
J.~Li, H.~Lee, and K.~J. Vahala, ``{Microwave synthesizer using an on-chip
  Brillouin oscillator},'' \emph{Nature Communications}, vol.~4, pp. 1--7, Jun.
  2013.

\bibitem{RevModPhys.77.633}
M.~Fleischhauer, A.~Imamoglu, and J.~P. Marangos, ``Electromagnetically induced
  transparency: Optics in coherent media,'' \emph{Rev. Mod. Phys.}, vol.~77,
  pp. 633--673, Jul 2005.

\bibitem{Fleischhauer2000}
\BIBentryALTinterwordspacing
M.~Fleischhauer, S.~Yelin, and M.~Lukin, ``How to trap photons? storing
  single-photon quantum states in collective atomic excitations,'' \emph{Optics
  Communications}, vol. 179, no. 1��, pp. 395 -- 410, 2000.
  \url{http://www.sciencedirect.com/science/article/pii/S0030401899006793}
\BIBentrySTDinterwordspacing

\bibitem{PhysRevLett.107.133601}
\BIBentryALTinterwordspacing
V.~Fiore, Y.~Yang, M.~C. Kuzyk, R.~Barbour, L.~Tian, and H.~Wang, ``Storing
  optical information as a mechanical excitation in a silica optomechanical
  resonator,'' \emph{Phys. Rev. Lett.}, vol. 107, p. 133601, Sep 2011.
  \url{http://link.aps.org/doi/10.1103/PhysRevLett.107.133601}
\BIBentrySTDinterwordspacing

\bibitem{PhysRevLett.92.083901}
\BIBentryALTinterwordspacing
M.~F. Yanik and S.~Fan, ``Stopping light all optically,'' \emph{Phys. Rev.
  Lett.}, vol.~92, p. 083901, Feb 2004.
  \url{http://link.aps.org/doi/10.1103/PhysRevLett.92.083901}
\BIBentrySTDinterwordspacing

\bibitem{Rakich:2012et}
P.~Rakich, C.~Reinke, R.~Camacho, P.~Davids, and Z.~Wang, ``{Giant Enhancement
  of Stimulated Brillouin Scattering in the Subwavelength Limit},''
  \emph{Physical Review X}, vol.~2, no.~1, p. 011008, Jan. 2012.

\bibitem{turik2006negative}
A.~Turik, A.~Yesis, and L.~Reznitchenko, ``Negative longitudinal
  electrostriction in polycrystalline ferroelectrics: a nonlinear approach,''
  \emph{Journal of Physics: Condensed Matter}, vol.~18, no.~20, p. 4839, 2006.

\bibitem{Ding2014}
Y.~Ding, L.~Chen, and S.~Shen, ``{Slow and fast light based on SBS with the
  spectrum tailoring},'' \emph{Optik - International Journal for Light and
  Electron Optics}, vol. 125, pp. 2181--2184, 2014.

\end{thebibliography}


\clearpage

\title{Supplementary Material: Non-Reciprocal Brillouin Scattering Induced Transparency}

\author{JunHwan Kim$^1$, Mark C. Kuzyk$^2$, Kewen Han$^1$, \\Hailin Wang$^2$, Gaurav Bahl$^{1\ast}$\\
	\\
	\footnotesize{$^1$Mechanical Science and Engineering, University of Illinois at Urbana-Champaign,}\\
	\footnotesize{Urbana, Illinois, USA}\\
	\footnotesize{$^2$Department of Physics, University of Oregon}\\
	\footnotesize{Eugene, Oregon, USA}\\
	\footnotesize{$^\ast$To whom correspondence should be addressed; E-mail: bahl@illinois.edu.}
}
\date{}
\maketitle

\section{Brief Review of Spontaneous and Stimulated Brillouin Scattering}
Spontaneous Brillouin scattering occurs when light interacts with the refractive index perturbations in a material caused by the presence of an acoustic wave. The spatio-temporal beat of the incident and scattered light fields then create a periodic variation in refractive index through electrostriction pressure. In the case that light is scattered to lower frequencies (i.e. Stokes scattering), the electrostriction pressure imparts energy to the sound wave. On the other hand, for anti-Stokes scattering, energy is removed from the sound wave leading to cooling and linewidth broadening \cite{Bahl:2012jm}. When sufficient input laser power is provided, the Stokes scattering process can overcome all intrinsic losses, resulting in the formation of a Brillouin laser through Stimulated Brillouin Scattering (SBS).

\begin{figure}[ht!]
	\begin{adjustwidth}{-1in}{-1in}
		\includegraphics[scale=0.8]{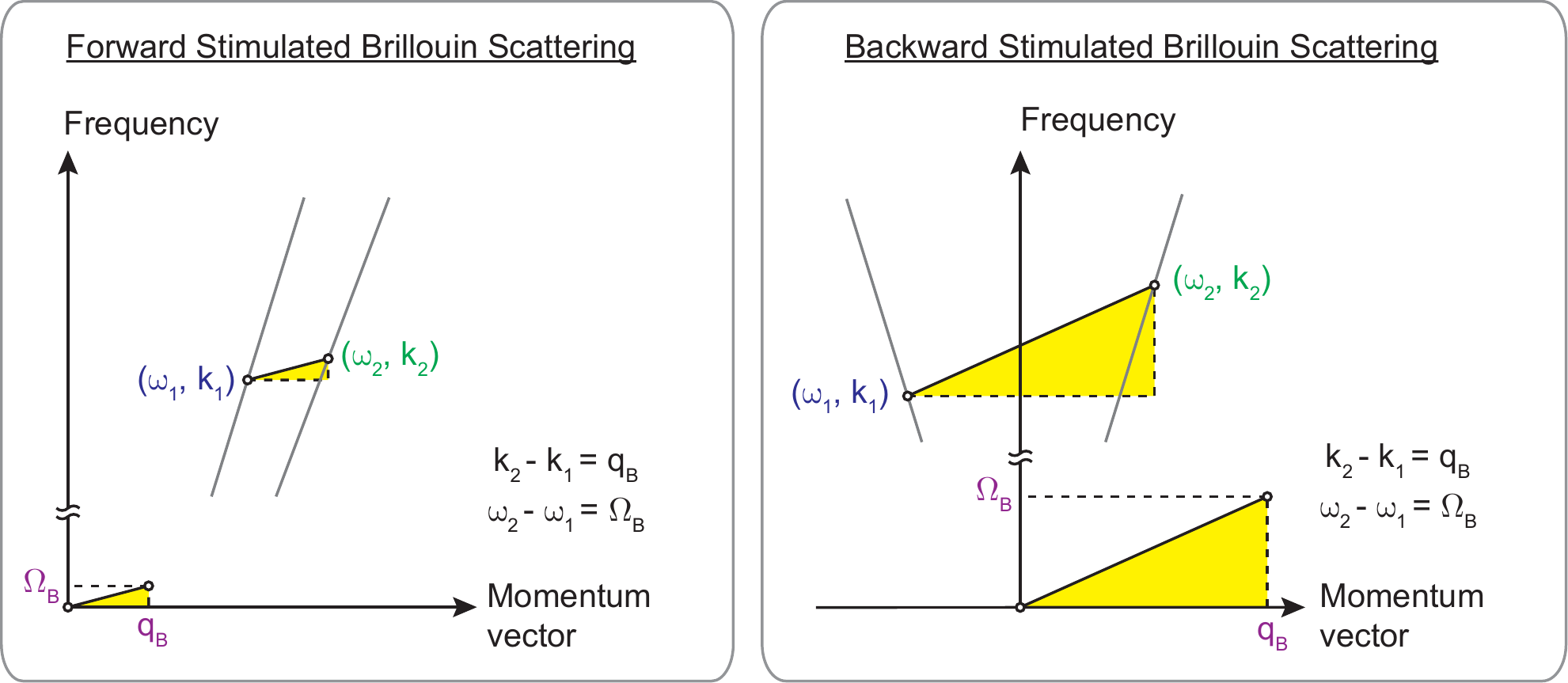}
		\caption{Dispersion diagram and wave vector matching conditions. When the control laser is parked at high frequency optical mode, ($\omega_2$, $k_2$), Stokes scattered light is generated at low frequency optical mode, ($\omega_1$, $k_1$). For Stokes scattering, the frequencies and wave vectors of the two optical modes and an intermediate acoustic mode must satisfy the condition, $k_2-k_1=q_B$ and $\omega_2-\omega_1=\Omega_B$. (left) Forward-SBS system. (right) Backward-SBS system.}
		\label{SBS}
	\end{adjustwidth}
\end{figure}

As shown in Fig.~\ref{SBS}, for both forward scattering and back-scattering, very specific energy conservation and momentum conservation (i.e. phase matching) requirements must be satisfied for Brillouin scattering processes to take place. In a back-scattering SBS system, the optical fields propagate in opposite directions. Since the optical k-vectors are nearly identical, the acoustic momentum vector is about double the length of the optical such that $|q_B| = |k_1| + |k_2|$. This implies the generation of acoustic waves in the tens of GHz frequency regime, depending on the refractive index and speed of sound in the material. In a forward-SBS system, the frequencies of the incident and scattered light are nearly identical. Hence, the acoustic frequency is typically in the sub-GHz range and the phonon lifetimes are significantly longer. The lower frequency is also implied through the necessarily short acoustic momentum vector in the forward case.

\section{Analytical Formulation for Brillouin Scattering Induced Transparency}

\begin{figure}[ht!]
	\includegraphics[trim = 0in 4.1in 0in 0in, clip, scale=0.65]{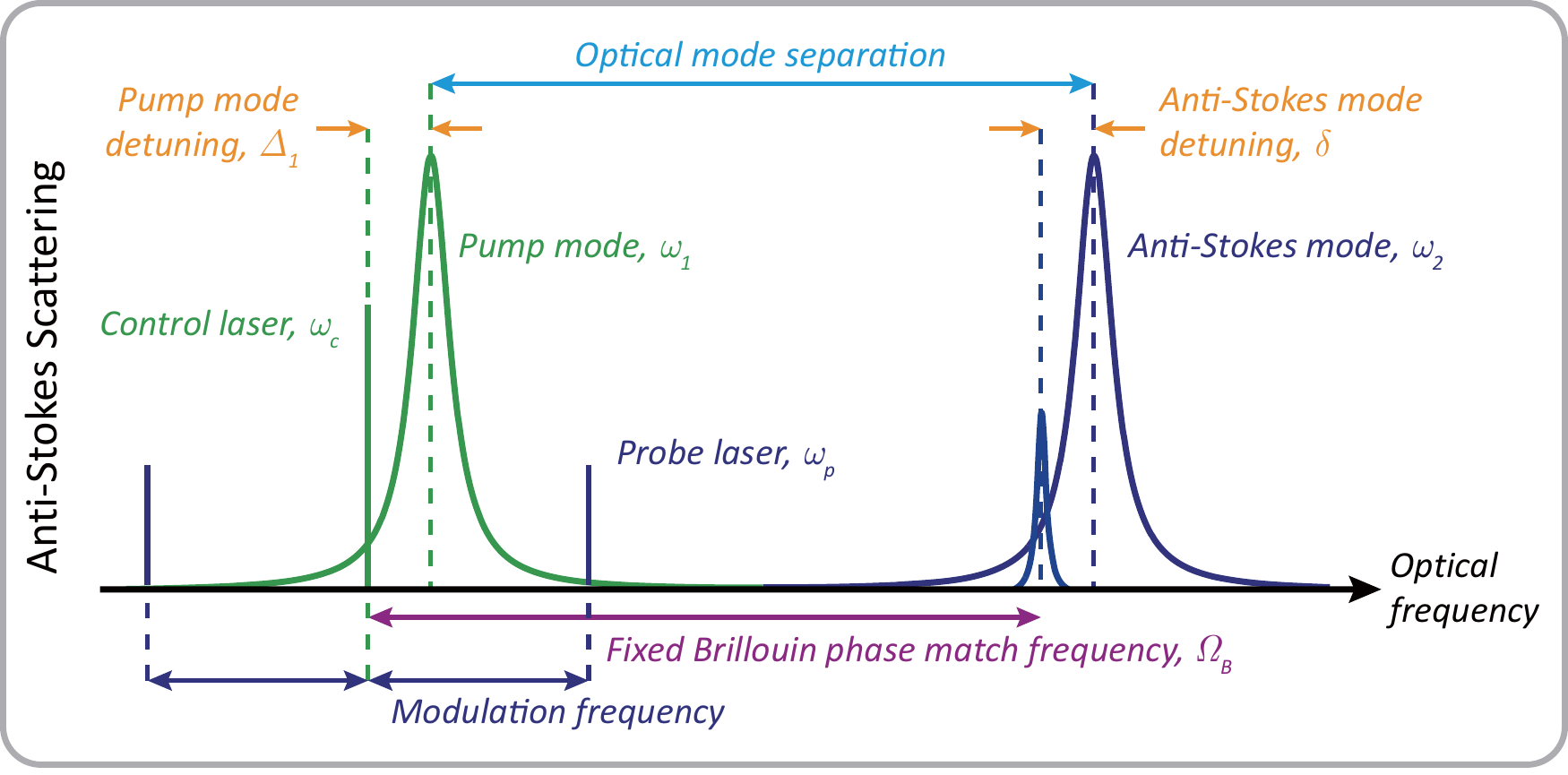}
	\vspace{-5pt}
	\caption{Optical frequency relationship of the coupled triplet system for transparency experiment using anti-Stokes scattering. The pump mode is at lower frequency than the anti-Stokes mode. When the modulation frequency sweeps over the fixed Brillouin phase match frequency, $\Omega_B$, the transparency is observed in the probe response.}
	\label{fig:Optical-frequency-relationship}
	
\end{figure}

For the analytical description of Brillouin scattering induced transparency (BSIT), we adopt the mathematical formalism established by Agarwal and Jha~\cite{PhysRevA.88.013815}. 
The intracavity fields representing the pump/control laser, anti-Stokes shifted probe, and acoustic displacement can be described using the following three coupled rate equations. 

\begin{equation}
\begin{aligned}
\dot{a_{1}} & = -\kappa_{1}a_{1}-i\Delta_{1}a_{1}-i\beta^{*}u^{*}a_{2}e^{-i\delta t}\\
\dot{a_{2}} & = -\kappa_{2}a_{2}-i\Delta_{2}a_{2}-i\beta ua_{1}e^{i\delta t}\\
\dot{u} & = -\Gamma_B u-i\Delta_{B}u-i\beta^{*}a_{1}^{*}a_{2}e^{-i\delta t}\label{AS_setup}
\end{aligned}
\end{equation}

\begin{equation}
\begin{aligned}
\delta & = \omega_{2}-\omega_{1}-\Omega_{B}		 \\
\Delta_{1} & = \omega_1-\omega_c		 \\
\Delta_{2} & = \omega_2-\omega_p		 \\
\Delta_{B} & = \Omega_B-(\omega_p-\omega_c)		\label{AS_detuning}
\end{aligned}
\end{equation}

where $a_{1}$, $a_{2}$, and $u$ are the slowly varying phasor amplitudes of intracavity
control field, scattered light field and mechanical displacement
respectively, $\kappa_{1}$ and $\kappa_2$ are optical loss rates of pump mode and anti-Stokes mode respectively, $\Gamma_B$ is acoustic loss rate, and $\beta$ is the coupling coefficient accounting for modal overlap and Brillouin gain in the material. The frequencies $\omega_1$, $\omega_2$, and $\Omega_B$ represent the pump optical resonance, anti-Stokes optical resonance, Brillouin acoustic resonance, while $\omega_c$, and $\omega_p$ represent the control laser field and probe laser field respectively. $\delta$, $\Delta_1$, $\Delta_2$, and $\Delta_B$ are the detuning parameters.

Details on the evaluation of detuning parameters and the coupling parameter $\beta$ are provided in \cite{PhysRevA.88.013815}. For phase matching, the frequency relationship $\omega_2 = \omega_1 + \Omega_B$ must be satisfied. Momentum matching is implicit in the complex phasors that represent the fields.

For the induced transparency experiment, we analyze this system at steady state, thus setting all derivatives to zero. Additional intracavity control field $f_{c}$ and probe field $f_{p}$ terms are added on the right-hand-side as shown in eqns.~\ref{AS_setup2}. For further simplification, we assume that the optical loss rates $\kappa_1$ and $\kappa_2$ are nearly identical (new symbol $\kappa$). Finally, the non-depleted pump field approximation eliminates the coupling term from the first equation. We then obtain the simplified system:
\begin{equation}
\begin{aligned}
0 & = -\gamma_1a_{1}+f_{c}	 \\
0 & = -\gamma_2a_{2}-i\beta ua_{1}e^{i\delta t}+f_{p}	 \\
0 & = -\gamma_Bu-i\beta^{*}a_{1}^{*}a_{2}e^{-i\delta t}		\label{AS_setup2}
\end{aligned}
\end{equation}
where
\begin{equation}
\begin{aligned}
\gamma_1 & = \kappa+i\Delta_1	 \\
\gamma_2 & = \kappa+i\Delta_2	 \\
\gamma_B & = \Gamma_B+i\Delta_B		\label{AS_gamma}
\end{aligned}
\end{equation}

\begin{figure}[t!]
	\includegraphics[trim = 0in 6.2in 0in 0in, clip, scale=0.82]{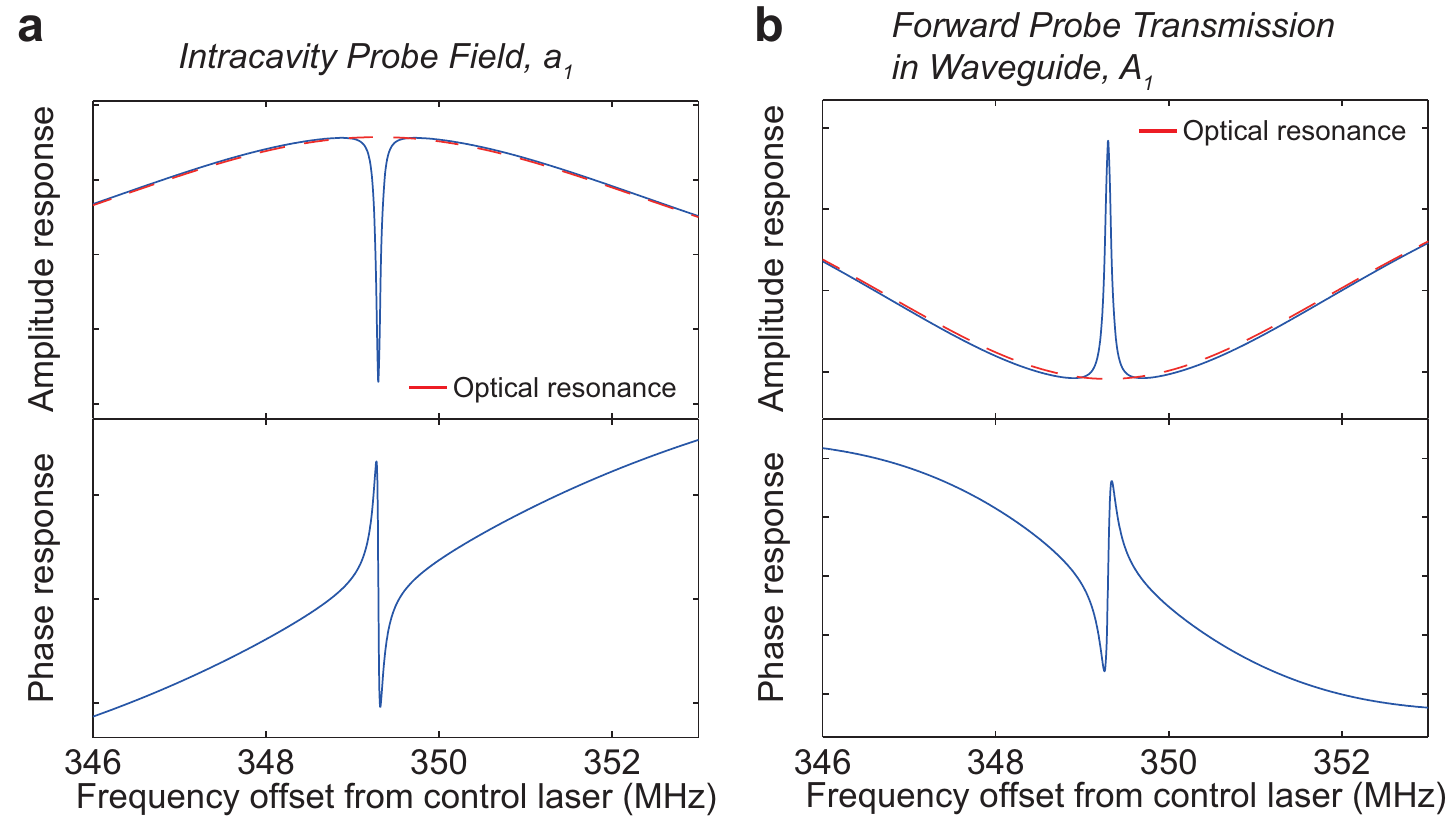}
	\vspace{-20pt}
	\caption{\textbf{Theoretical prediction of amplitude and phase responses for induced transparency.} \textbf{a.} Probe field when measured inside the cavity. \textbf{b.} Probe field transmitted and measured at the photodetector. The phase response of intracavity probe field is inverted as the light evanescently couples back to the waveguide and mix with the part of probe field that was reflected from the cavity.}
	\label{antiStokesFields}
	
\end{figure}

The system of eqns.~\ref{AS_setup2} can then be solved to produce the steady state amplitudes of the fields:
\begin{align}
a_{1} & = \frac{f_{c}}{\gamma_1}\label{AS_a_1}\\
a_{2} & = \frac{f_{p}\gamma_B}{\gamma_2\gamma_B+|\beta|^{2}|a_{1}|^{2}}\label{AS_a_2}\\
u & = \frac{-i\beta^{*}a_{1}^{*}a_{2}e^{-i\delta t}}{\gamma_B}\label{AS_u}
\end{align}
The control laser (eq.~\ref{AS_a_1}) excites the system, while the probe laser, described by eq.~\ref{AS_a_2}, sweeps through the anti-Stokes optical mode of interest and experiences the induced transparency. The intracavity probe field transfer function is illustrated in Fig.~\ref{antiStokesFields}a.

Note that the phase response of the probe within the cavity is in agreement with the results from previous SBS demonstrations, that is to say an anti-Stokes probe experiences a fast light response. 
However, the opposite result (slow light) is observed when monitoring the probe field in the waveguide ($A_2 = R F_p + i T a_2$). Here, the input probe laser field is related to input intracavity field as $F_p = - i f_p/T$~\cite{GorodetskyOptimal}, while $R$ and $T$ are the reflection and transmission coefficients at the coupler. As shown in Fig.~\ref{antiStokesFields}b, we observe a slow light behavior for the probe when measured in the waveguide.

\vspace{12pt}

\section{Analytical Formulation for Induced Absorption}

\begin{figure}[t!]
	\includegraphics[trim = 0in 0in 0in 3.5in, clip, scale=0.65]{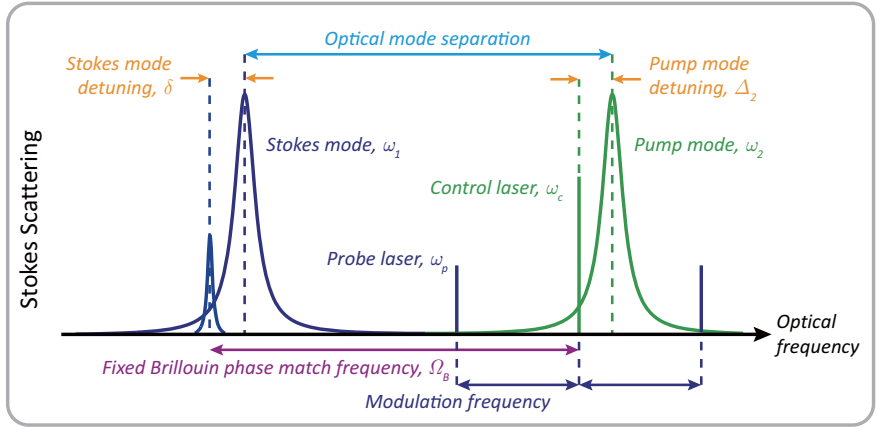}
	\vspace{-35pt}
	\caption{Optical frequency relationship of the coupled triplet system for absorption experiment using Stokes scattering. Opposite to anti-Stokes scattering case, the role of the pump and scattered modes is switched. Also, the probe laser sweeps in the opposite direction from higher to lower frequency.}
	\label{fig:Stokes_Optical-frequency-relationship}
\end{figure}

\begin{figure}[ht!]
	\includegraphics[trim = 0in 2.2in 0in 3.9in, clip, scale=0.82]{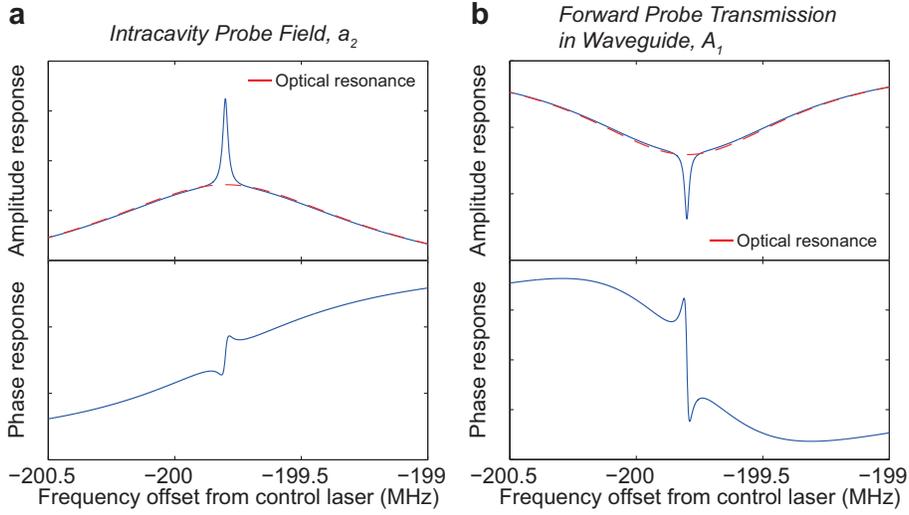}
	\vspace{-20pt}
	\caption{\textbf{Theoretical prediction of amplitude and phase responses for induced absorption.} \textbf{a.} Probe field when measured inside the cavity. \textbf{b.} Probe field transmitted and measured at the photodetector. The phase response of intracavity probe field is inverted as the light evanescently couples back to the waveguide and mix with the part of probe field that was reflected from the cavity.}
	\label{StokesFields}
	
\end{figure}

To understand Brillouin scattering induced absorption, we must consider the process with a Stokes probe. Here, we employ the same set of equations used for induced transparency (eq.~\ref{AS_setup}), except that we reverse the roles of the control and probe lasers. In other words, subscript 1 refers to the Stokes probe while subscript 2 refers to the control field. We can then rewrite the simplified system equations as:
\begin{equation}
\begin{aligned}
0 & = -\gamma_1a_{1}-i\beta^{*}u^{*}a_{2}e^{-i\delta t}+f_{p}\\
0 & = -\gamma_2a_{2}+f_{c}\\
0 & = -\gamma_Bu-i\beta^{*}a_{1}^{*}a_{2}e^{-i\delta t}	\label{S_setup2}
\end{aligned}
\end{equation}
where
\begin{equation}
\begin{aligned}
\gamma_1 & = \kappa + i\Delta_1\\
\gamma_2 & = \kappa + i\Delta_2\\
\gamma_B & = \Gamma_B + i\Delta_B
\end{aligned}
\end{equation}
The frequency matching relationship between the fields is unchanged i.e. $\omega_2 = \omega_1 + \Omega_B$. However, as illustrated in Fig.~\ref{fig:Stokes_Optical-frequency-relationship}, the detuning parameters are modified on account of the interchanged control and probe designations.
\begin{equation}
\begin{aligned}
\Delta_1 & = \omega_1 - \omega_p\\
\Delta_2 & = \omega_2 - \omega_c\\
\Delta_B & = \Omega_B - (\omega_c - \omega_p)
\end{aligned}
\end{equation}
Upon solving system \ref{S_setup2}, the intracavity probe field is described as 
\begin{align}
a_{1} = \frac{f_{p}\gamma_B^*}{\gamma_1\gamma_B^*-|\beta|^{2}|a_{2}|^{2}}\label{S_a_1}
\end{align}
As before, the forward probe transmission in the waveguide $A_1$ is described as
\begin{align}
A_1 = R F_p + i T a_1 ~.
\end{align}
Again, we note an inversion of phase response when the probe field exits the resonator and mixes with the reflected input probe that did not couple to the resonator (Fig.~\ref{StokesFields}).

\vspace{12pt}

\section{Experimental Details}

\begin{figure}[t!]
	\includegraphics[scale=0.6]{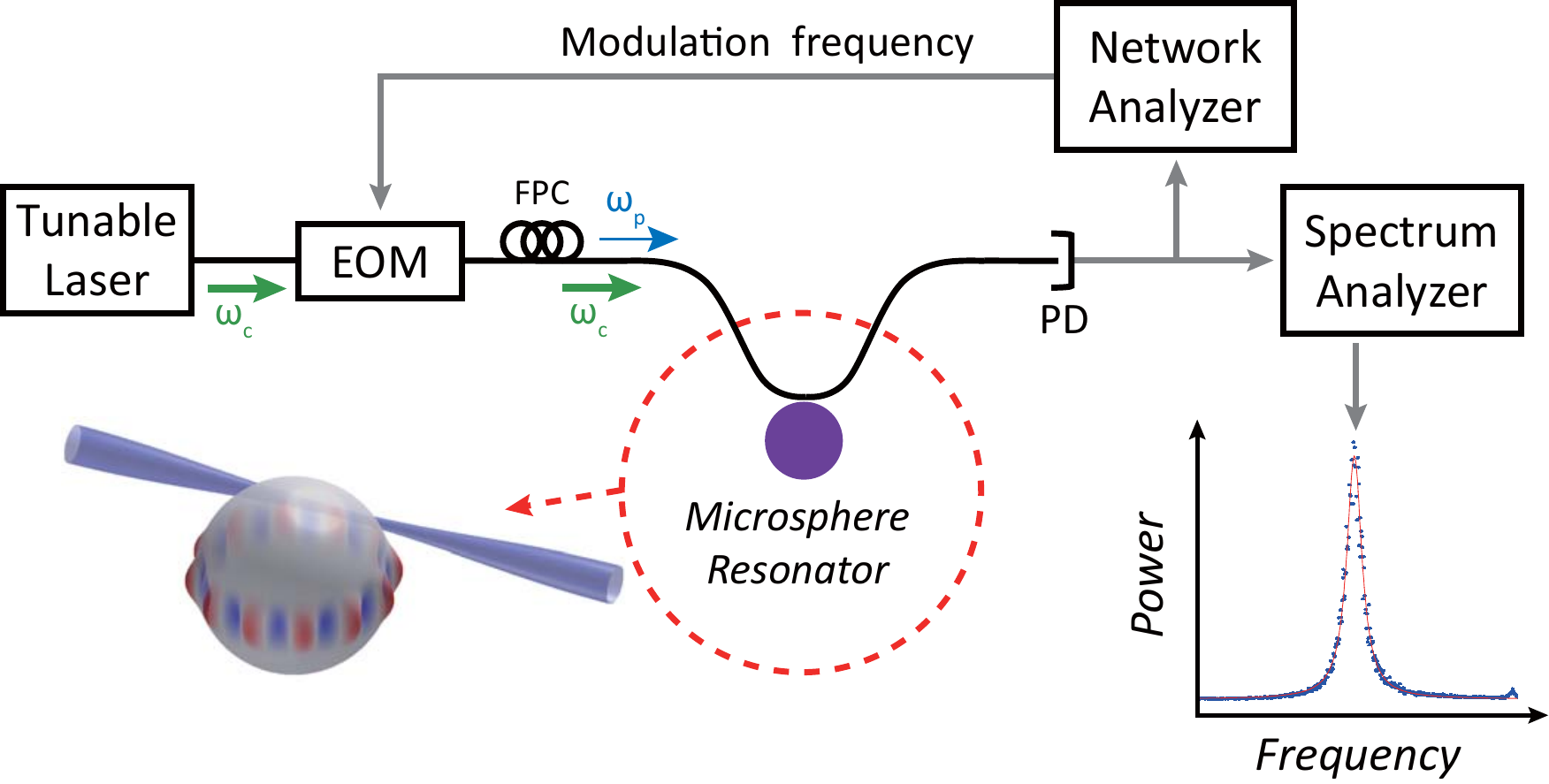}
	
	\caption{Experimental setup for BSIT. Fiber coupled tunable diode laser sends the control
		laser singal through a tapered waveguide. Network analyzer provides
		probe frequency to the electro-optic modulator which creates a probe
		signal from the control laser. Photodetector (PD) monitors the signal transmission
		at the waveguide output and the electronic signal is analyzed using
		both network analyzer and electrical spectrum analyzer. FPC is fiber polarization controller.}
	\label{setup}
	
\end{figure}

\begin{figure}[t!]
	\includegraphics[scale=0.8]{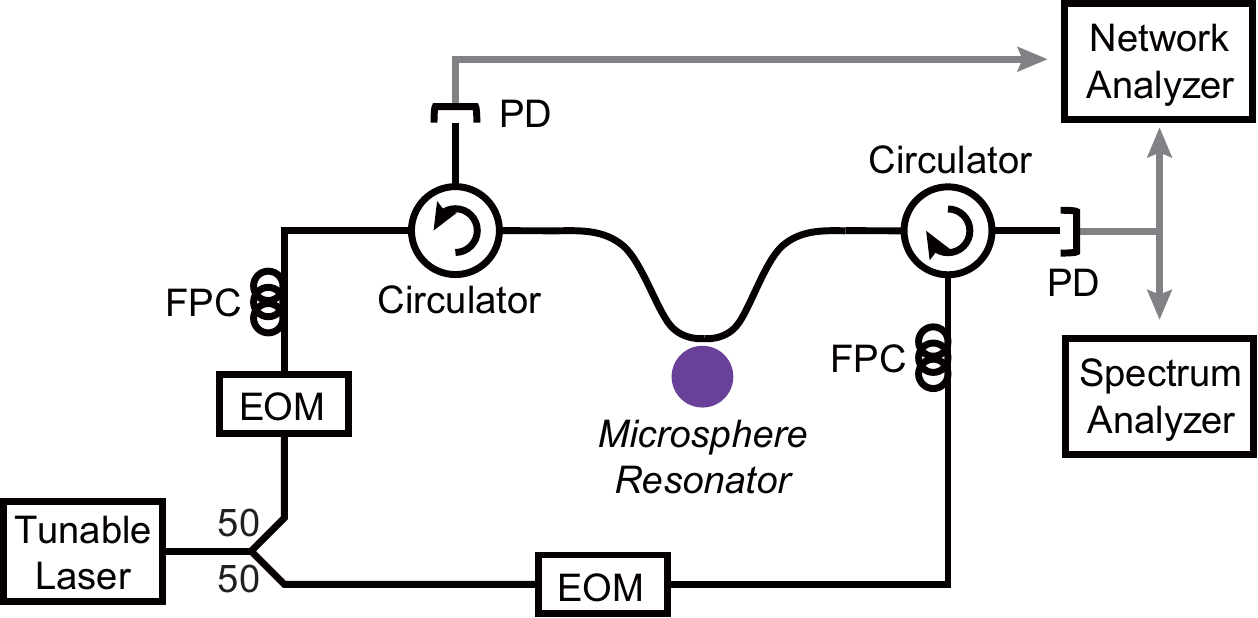}
	
	\caption{Experimental setup used for the non-reciprocity experiment. Independent probe signal can be sent in the forward and backward directions by controlling the two EOMs.}
	\label{isolationSetup}
	
\end{figure}

Fig.~\ref{setup} shows the experimental setup used in this work.
An ultra-high-Q silica microsphere resonator is evanescently coupled to a tapered silica fiber waveguide that provides an interface to the intracavity control and probe light fields. 
A tunable diode laser with a center wavelength of 1550 nm provides the control field to the pump mode, while the probe signal is generated using an electro-optic modulator (EOM).
While the EOM creates two probe sidebands relative to the control laser, only one sideband is matched to the anti-Stokes optical mode of the device. The other (Stokes) sideband passes through the system unhindered at constant amplitude. The probe sideband spacing is determined by a modulation signal input to the EOM that is generated by a network analyzer.
The transmitted optical signal at the output of the waveguide is received by a highspeed photodetector, which results in an electronic output beat note between the control and probe optical signals. This electronic output is analyzed by an electrical spectrum analyzer as well as the network analyzer for probe transfer function analysis. 
When the higher energy optical mode is pumped above threshold (in the absence of a probe), the electronic beat note can be used to distinguish the Brillouin scattering~\cite{Bahl:2012jm,Bahl:2011cf} from radiation pressure induced optomechanical oscillation \cite{PhysRevLett.95.033901}.

The experimental setup used for measuring non-reciprocal transmission in forward and backward directions is shown in Fig.~\ref{isolationSetup}. The control signal in the backward direction is carefully suppressed by biasing the EOM at a transmission null.

\section{Comparison Against Other SBS Systems}

The figure-of-merit reported in SBS slow light systems is the product of group delay and bandwidth \cite{Song:2005vp,Okawachi:2005cw,Thevenaz:2008vw,4139648} as it represents the information storage capacity for the system. We must note, however, that both laser power and device size are engineering-related constraints that must also be budgeted wisely. This is critically important in on-chip devices where the laser power budgets are extremely small and kilometer-long waveguides (as in fiber SBS systems) are simply impractical.

Thus, in order to make a comparison of the relative efficiency of our resonator approach against various other SBS based systems, we divide the achieved delay-bandwidth by the control laser power and device size through ${\tau \Gamma_B}/{I_c L}$, where $\tau$ is group delay, $\Gamma_B$ is bandwidth for the delay, $I_c$ is the control laser input power, and $L$ is opto-acoustic interaction length. In the case of linear waveguides, this length is the total waveguide length \cite{Song:2005vp,Okawachi:2005cw,Thevenaz:2008vw,4139648} while in resonators it is fair to use the resonator circumference as a linear measure of the optical path. Indeed, resonators have the advantage of the high optical and acoustic finesse compared against linear systems.

We compare our experimental data against previous SBS slow light reports in Table~\ref{FOM_table}. It is seen that our system provides a $\tau \Gamma_B$ product that is comparable to all previous demonstrations (order-of-magnitude is 1). Considering size and power, however, our system provides a $\tau \Gamma_B$ product with 5 orders-of-magnitude lower power $\times$ length product than the next nearest system. This enormous engineering advantage could make our system particularly well-suited for compact, ultralow power, on-chip SBS slow light systems that are otherwise impractical.

\begin{landscape}
	\begin{centering}
		\begin{table}[ht!]
			\vspace{50pt}
			\caption{This table compares the figures-of-merit for various SBS slow light systems. For each publication, only the results with highest delay $\times$ bandwidth product ($\tau \Gamma_B$) are presented. Our system is on par with other SBS slow light systems. However, the power- and size- normalized delay $\times$ bandwidth shows that our resonator system can provide comparable $\tau \Gamma_B$ with $10^5$ times lower power and size when compared against the next nearest prior result.}
			\vspace{10pt}
			
			\begin{tabular}{| >{\raggedright\arraybackslash}m{3.5cm} | >{\raggedright\arraybackslash}m{2cm} | >{\raggedright\arraybackslash}m{2cm} | >{\raggedright\arraybackslash}m{2cm} | >{\raggedright\arraybackslash}m{2cm} | >{\raggedright\arraybackslash}m{2cm} | >{\raggedright\arraybackslash}m{2.8cm} |}
				
				\hline
				
				Author & Highest group delay, $\tau$ ($\mu$s) & Bandwidth, $\Gamma_B$ (MHz) & Control laser power, $I_c$ (mW) & Interaction length$^*$, $L$ (km) & Delay $\times$ Bandwidth product, $\tau \Gamma_B$ ($\mu$s-MHz) & Power and size normalized Delay $\times$ Bandwidth, $\dfrac{\tau \Gamma_B}{I_cL}$ ($W^{-1}m^{-1}$) \\ \hline
				This work (Slow light) & 110 & 0.017 & 1 & 4.7$\times 10^{-7}$ & 1.87 & 3.98$\times 10^{6}$ \\ \hline
				R. Pant~\cite{Pant2012} & 0.023 & 40$^\dagger$ & 300 & 7$\times 10^{-5}$ & 0.92 & 4.38$\times 10^{1}$ \\ \hline
				K. Y. Song~\cite{Song:2005vp} & 0.018 & 13.33$^\dagger$ & 0.012 & 6.7$\times 10^{0}$ & 0.24 & 2.99 \\ \hline
				L. Yi~\cite{4139648} & 0.00052 & 1250 & 200 & 1.25$\times 10^{1}$ & 0.65 & 2.6$\times 10^{-1}$ \\ \hline
				H. Ju~\cite{6582552} & 0.04 & 11.24$^\dagger$ & 20 & 2$\times 10^{0}$ & 0.45 & 1.12$\times 10^{-2}$ \\ \hline
				Y. Okawachi~\cite{Okawachi:2005cw} & 0.02 & 66.67$^\dagger$ & 250 & 5$\times 10^{-1}$ & 0.4 & 1.07$\times 10^{-2}$ \\ \hline
				Y. Ding~\cite{Ding2014} & 0.024 & 50$^\dagger$ & 16 & 5$\times 10^{1}$ & 1.2 & 1.5$\times 10^{-3}$ \\ \hline

			\end{tabular}
			
			\vspace{12pt}
			$^*$ Waveguide length or resonator circumference.
			
			$^\dagger$ Bandwidth calculated from the pulse width.
			\label{FOM_table}
		\end{table}
	\end{centering}
\end{landscape}

\section{Strong Coupling Regime}

When the coupling rate is comparable to the optical loss rate $|\beta||a_1| \approx \kappa $, the system enters the strong coupling regime~\cite{RevModPhys.77.633}. In the strong coupling regime, we observe a mode split in which an optical resonance at the original frequency is completely removed. Instead, we have effectively two new optical resonances which can be tuned using the coupling rate.

\begin{figure}[ht!]
	\begin{adjustwidth}{-1in}{-1in}
		\includegraphics[scale=0.78]{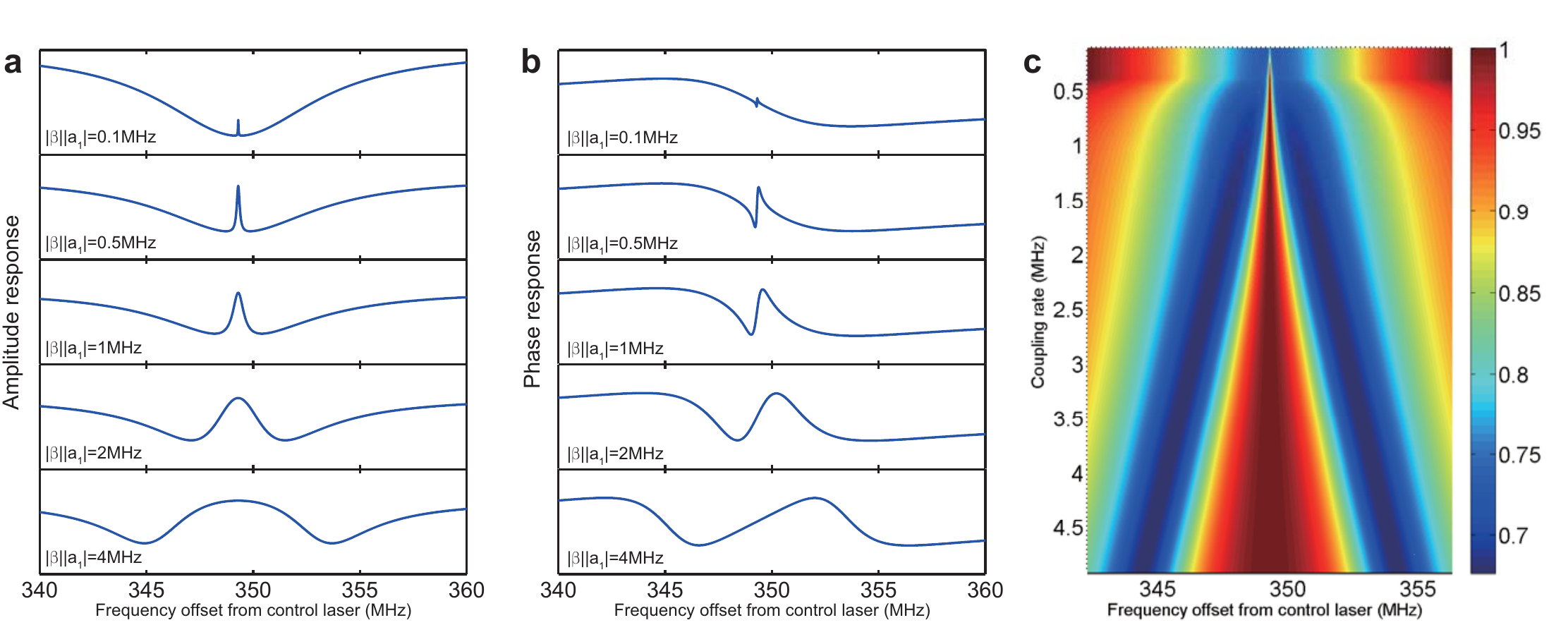}
		\caption{\textbf{Analytical simulations of tuning transparency using control laser power.} \textbf{a.} Increasing the control laser power increases the coupling rate, $|\beta||a_1|$, and the coupling rate increases the amplitude of the transparency peak. Past the strong coupling regime, the mode splitting occurs. \textbf{b.} The change in phase response with increasing control laser power. In a. and b., the first row corresponds to our experimental result shown in manuscript Fig.~2. \textbf{c.} Spectrogram of normalized amplitude response. Color bar on the right represents the absorbed optical power. The degree of mode split increases with increasing control laser power.}
		\label{mode}
	\end{adjustwidth}
\end{figure}

Experimentally, the coupling coefficient $\beta$ can be held constant while the control field $a_1$ is used to manipulate the coupling rate. In Fig.~\ref{mode}a, the progression of the mode split with increasing control laser power is calculated. We note that the strong coupling regime could be reached with only 40 $\mu$W of dropped input optical power. Here, optical loss rate $\kappa$=4.4 MHz and acoustic loss rate $\Gamma_B$=16.9 kHz are extracted from the experimental data.

\end{document}